\documentstyle[12pt,aaspp4,flushrt]{article}

\input epsf
\def\lax    {{_{\textstyle <}\atop^{\textstyle \sim}}}

\begin{document}

\noindent Astronomy Letters, 1996, V.22, N.10 (in press)

\bigskip

{\Large \bf Scattering of X-ray emission lines by the neutral and molecular
hydrogen in the Sun's atmosphere and in the vicinity of active galactic nuclei
and compact sources}

\bigskip

\centerline{R.Sunyaev$^{1,2}$ E.Churazov$^1$}

\bigskip

\centerline{$^1$Space Research Institute (IKI), Profsouznaya 84/32, Moscow 117810, Russia}
\centerline{$^2$ MPI fur Astrophysik,
Karl-Schwarzschild-Strasse 1, 86740 Garching bei Munchen, Germany}


\bigskip

\pagebreak

\section*{Abstract}
In many astrophysical objects an account for
scattering of the  X-ray fluorescent and resonance lines of iron ions by  the
electrons, bound in atomic and molecular hydrogen is important. Specific
distortions of the scattered line profile 
spectrum appear in comparison with scattering by free electrons. Analysis of
scattered line spectra may provide information on the ionization state,
helium abundance and geometry of scattering media in the vicinity of
AGNs, compact X-ray sources and on the surface of X-ray flaring stars
and the Sun.
\clearpage

\section{Introduction}
Compton scattering of X-rays by  free electrons is known to play an
important role in a number of astrophysical objects: early Universe, hot
gas in clusters of galaxies, accretion disks around neutron stars and black
holes, plasma clouds surrounding compact X-ray sources. In many cases X-rays
are scattered by neutral gas rather than free electrons. One can mention 
the reflection of X-rays by the solar photosphere
(X-rays are produced by the flares above the solar surface and a considerable
fraction of them will be reflected by the photosphere, having low degrees of
ionization $\frac{e}{H}\le 10^{-3}$); reflection of X-rays by the
photospheres of cold flaring stars (T-Tauri, late stars coronas); outer
regions of the extended accretion disks in binary systems, having low degree
of ionization; molecular tori around central objects in the active galactic
nuclei and QSOs. Compact X-ray sources, embedded into the dense molecular
clouds are of special interest. Discovery of the giant molecular cloud in
the direction of the source 1E1740.7--2942 (Bally and Leventhal 1991)
indicates that the source is possibly surrounded by the dense molecular gas.
Provided the strong variability of the compact source, observations of
scattered component (leaving the cloud with substantial time lag) are
possible, allowing one to study the parameters of the molecular cloud by
means of X-ray astronomy methods.

Sunyaev, Markevitch and Pavlinsky (1993) have noted that strong
concentration of the molecular hydrogen in the giant clouds in the Galactic
Center region should lead to the appearance of the diffuse emission, related
to the scattering of the compact X-ray sources and Sgr A* by the molecular
hydrogen in these clouds. It was also noted, that detection of a strong
fluorescent line of neutral iron ($h\nu=$ 6.4 keV) and characteristic
spectrum of scattered component may provide unique information on the
activity of this region during last several hundreds years. Moreover,
scattering of the emission from the nucleus of the Galaxy by the neutral and
molecular cloud of the whole Galaxy should lead to the appearance of the
diffuse component, which contains the information of the activity of the
nucleus during past tens of thousands of years. Recent ASCA observations of the
diffuse emission of the Galactic Center region (Koyama, 1994) gave new
information in support of this hypothesis. Along with emission lines from
the hydrogen- and helium-like ions of heavy elements, the powerful iron
fluorescent  line ($K_{\alpha}$) was detected, which surface brightness
distribution roughly follows the distribution of giant molecular
clouds. In particular the molecular complex Sgr B2 was found to be especially
bright in this line.

        Bolometric observations with high energy resolution (better than 10
eV) of ASTRO-E, Bragg spectrometer (resolution $\sim$ eV) of
Spectrum-X-Gamma and gratings of XMM and AXAF
can verify another prediction -- specific shape of the low energy wing of
the lines formed due to scattering by atomic or molecular hydrogen. The
shape of 
the low energy wing is related to the recoil effect and contains the
information on the
geometry of scattering. We will show below that recoil effect on
bound electrons differs substantially from that on the free
electrons. Thus, the spectral shape of the scattered line can be used as
diagnostic of the ionization balance in the scattering medial in distant
AGNs and QSOs. Moreover studying shapes of the scattered lines opens the 
principal possibility of helium abundance determination at the Sun's surface
and in molecular clouds.

Major effects are the {\it smearing of back scattering peak and appearance
of the energy gap} -- lack of photons in the interval from initial photon
energy $h\nu_0$ down to $h\nu_0-E$, where $E$ is the excitation energy of
the first levels above ground level ($E=$ 10.2 eV for atomic hydrogen,
$E\approx$ 20 eV for neutral helium, $E=$ 40.8 eV for He II ion, $E\approx$
11 eV for molecular hydrogen). The possibility of the detection of Raman
satellites of X-ray emission lines is very attractive. The energies of these
lines are shifted by the excitation energy of various levels of the atoms of
hydrogen and helium. Of course with the current X-ray CCD and cooled solid
state detectors with an energy resolution about 100 eV one can expect detection
of effects only in the zone with $\Delta h\nu\sim 100$ eV. As is shown below
these effects are large enough.

\section{Scattering by hydrogen atoms}
        All astrophysicists, dealing with detectors of X and Gamma-ray
emission, know that Compton scattering takes place in the neutral gas or
solid matter of the detector (e.g. \cite{zom}). \cite{bs73} and \cite{bst74}
for the first time considered in detail 
the problem of heating the surface of the atmosphere of normal star in the
close binary system HZ Her/Her X-1 and mentioned the importance of
scattering and reflection of X-rays by the weakly ionized atmosphere of Hz
Her. They noted that for the photons with energy above a few keV the
recoil effect exceeds the ionization potential of hydrogen. Thus in the
first approximation scattering of such photons by neutral hydrogen does not
differ from scattering by free electrons. The same approximation was used by
\cite{bas78} and \cite{gf91}, considering the reflection of X-rays by the
stellar surface and cold accretion disk.

        The necessity of more detailed treatment arises because of the
appearance of the new generation of X-ray  instruments with  
the energy resolution $\sim$ few eV at the fluorescent line 6.4 keV, and
resonance lines of hydrogen-- and helium--like ions of iron. Below we
discuss the effects which can be observed by the instruments which are going
to be launched by the year 2000 (Spectrum-X-Gamma with Bragg spectrometer, AXAF
and XMM with gratings, ASTRO-E with X-ray bolometers). 

        Most of the relations describing the process of X-ray photon
scattering  by free or bound electrons are well known and can be found in
the textbooks and original papers. We  give some below
for completeness.

For the scattering of a photon by a free electron ($\gamma_1 + e_1=\gamma_2 +
e_2$) the frequency of the photon after scattering is
unambiguously defined by the geometry of scattering, due to energy and
momentum conservation laws:
\begin{eqnarray}        \label{recoil0}
\nu_2=\nu_1 \frac{1
-\frac{\vec{v}}{c}\vec{\Omega}_1}{1-\frac{\vec{v}}{c}\vec{\Omega}_2 +
\frac{h\nu_1}{\gamma m_ec^2}(1-\vec{\Omega}_1\vec{\Omega}_2)}
\end{eqnarray}
where $\vec{\Omega}_1, \nu_1$ and $\vec{\Omega}_2, \nu_2$ characterize the
direction and frequency of the photon before and after scattering, $\vec{v}$
is the velocity of the electron before scattering,
$\gamma=\frac{1}{\sqrt{1-\frac{v^2}{c^2}}}$,
$\vec{\Omega}_1\vec{\Omega}_2=cos\theta$, $\theta$ is the scattering angle
(see e.g. Kompaneetz, 1956, review of Pozdnyakov et al., 1982).

In the limit $\vec{v}=0$ and $\frac{h\nu_1}{m_ec^2} \ll 1$ change of the photon
energy is related to the recoil effect:
\begin{eqnarray}        \label{recoil1}
\nu_2=\nu_1 \left[1- \frac{h\nu_1}{m_ec^2}(1-cos\theta)\right]
\end{eqnarray}
In the limit $\frac{h\nu_1}{m_ec^2} \ll \frac{v}{c}$ Doppler effect
dominates:
\begin{eqnarray}        \label{doppler}
\nu_2=\nu_1
\frac{1-\frac{\vec{v}}{c}\vec{\Omega}_1}{1-\frac{\vec{v}}{c}\vec{\Omega}_2} 
\end{eqnarray}

        For the scattering by the electron, bound in a hydrogen atom,
additional factors complicate the process: finite binding energy of the
electron and motion of the electron in the atom. Because of discrete energy
levels of the electron in the exited state not all values of photon energy
change are possible. Because of the ``random'' motion of the electron within
the atom change of the photon energy is no longer unambiguous function 
of the geometry. For free electrons \cite{pss79} noted, that even at
rather low electron temperatures $kT_e\sim 1$ eV profile of a scattered line
apparently broadens  due to the Doppler effect (see e.g. Fig.7 in
\cite{pss82}). Note that in this example electron velocity was $v\approx
400$ km/s, the value not unusual for accretion disks in X-ray binary systems
and galactic nuclei. Characteristic velocity of the electron
in hydrogen atom is $v\sim \alpha c\sim 2000$ km/s ($1/\alpha= \hbar
c/e^2=137$ is the fine structure constant) and motion with such
velocity should significantly affect the amount of energy transferred to the
electron during the scattering. This ambiguity of energy transfer does not
violate 
conservation laws, since heavy nucleus can carry the necessary momentum, having
negligible kinetic energy.

Depending on the final state of the electron, the  scattering process can be
divided 
into three branches:
\begin{itemize}
\item[] {\bf A)} Rayleigh (coherent) scattering:
$\gamma_1+H=\gamma_2+H$. The energy of the photon remains the same, only
direction changes. The recoil effect is $\sim m_p/m_e$ times lower than in
the case of scattering by free electron. 
\item[] {\bf B)} Raman scattering:
$\gamma_1+H=\gamma_2+H(n,l)$, where $H(n,l)$ denotes one of the excited
states of the hydrogen atom. The energy of the photon decreases by the value
of excitation 
energy: $h\nu_2=h\nu_1-E_{n,l}$ -- Raman satellites of the line appear. 
\item[] {\bf C)} Compton scattering:
$\gamma_1+H=\gamma_2+e^-+p$, accompanied by ionization of atom. The energy 
of the photon decreases by the value of ionization potential and kinetic
energy of electron after scattering: $h\nu_2=h\nu_1-13.6 eV - E_e$. 
\end{itemize}
It is necessary to mention that in a non-relativistic limit sum over
differential crossections of these three branches is equal to the Thomson
differential crossection (see below).

Below we briefly discuss each of these three processes. More detailed
discussion on the process of scattering by hydrogen atom and references for
original publications can be found in e.g. \cite{ep70}.

The following notations will
be used 
throughout the paper: \\
$\nu_1, \nu_2, \vec{k_1}=\vec{\Omega}_1\frac{h\nu_1}{c},
\vec{k_2}=\vec{\Omega}_2\frac{h\nu_2}{c}$ are initial and final frequencies and
momenta of the photon, $\Delta\nu=\nu_1-\nu_2, \vec{q}=\vec{k_1}-\vec{k_2}$
are change of the frequency and momentum, $\vec{\chi}=\vec{q}/\hbar$,
$a=r_b/\hbar$, $r_b$ -- Bohr radius, $\theta$ is the
scattering angle, 
\begin{eqnarray}        \label{th}
\left(\frac{d\sigma}{d\Omega}\right)_{Th}=0.5\cdot r^2_e\cdot
(1+cos^2\theta)\left(\frac{\nu_2}{\nu_1}\right)^2 
\end{eqnarray}
is the differential crossection for scattering of the photon by a free
electron at 
rest in non-relativistic limit. This expression coincides with the familiar
expression $\left(\frac{d\sigma}{d\Omega}\right)_{Th}=0.5\cdot r^2_e\cdot
(1+cos^2\theta)$, except for the factor $\sim\frac{\nu_2}{\nu_1}\approx 1$,
which appears in the expression for crossection when a change of the photon
energy can't be neglected. We drop this factor in a few expressions below,
which is certainly valid for the considered range of photon energies.

The probability of photon scattering accompanied by the transition of electron
from the initial state $i$ to the final state $f$ can be calculated from 
perturbation theory. The crossection can be calculated as
$\frac{2\pi}{\hbar c} |X|^2 \rho\times \delta(\Delta E - h\nu)$ (e.g
\cite{hei}), where $\rho\approx\frac{k^2_2 d\Omega}{(2\pi \hbar c)^3}$ is
the density of the final states per unit energy,  $\Delta E$ is the change of
electron energy. In the first order perturbation
theory$X=H'_{if}=\int\psi^*_f H' \psi^*_i$ (e.g. \cite{llqm}, \cite{hei}).
$H'$ in 
the above expression stands for the term in Hamiltonian responsible for the
interaction of the electron and photon. In non-relativistic approximation
\begin{eqnarray}        \label{ham}
H' = e^2A^2/2mc^2-e\vec{p}\vec{A}/mc
\end{eqnarray}
(e.g. \cite{lltp,hei}), where  $\vec{A}$ is the vector-potential of
the electro-magnetic field,
$\vec{p}$ is the momentum of the electron. Scattering of the photon appears
in the first 
order perturbation theory for the term $e^2A^2/2mc^2$ in the expression
(\ref{ham}) and in the second order for the term $e\vec{p}\vec{A}/mc$.
In the case under consideration (scattering of photons with energy $\sim$ few
keV by 
light elements) the second term in the expression (\ref{ham}) can be neglected
(e.g. \cite{hei},\cite{ep70}). Substituting $\vec{A}$ with the plane wave
one gets the expression for the matrix element for transition $i\rightarrow
f$:
\begin{eqnarray}        \label{ham1}
H'_{if}=\frac{e^2}{mc^2}(\vec{e_1}\vec{e_2})\frac{2\pi\hbar^2c^2}{\sqrt{h\nu_1
h\nu_2}} \int \psi^*_f e^{i\vec{\chi}\vec{r}} \psi_i=
\frac{e^2}{mc^2}(\vec{e_1}\vec{e_2})\frac{2\pi\hbar^2c^2}{\sqrt{h\nu_1 h\nu_2}}\langle f | e^{i\vec{\chi}\vec{r}} | i \rangle
\end{eqnarray}
$\vec{e_1}$ and $\vec{e_2}$ in the above expression stand for the unit
vectors in the direction of polarization of the photon before and after
scattering. Thus final expression for the differential
crossection of photon scattering with a given change of frequency, direction
and polarization will be as follows:
\begin{eqnarray}        \label{seq}
\frac{d\sigma}{d\Omega dh\nu}=\left(\frac{e^2}{mc^2}\right)^2 \left(\frac{\nu_2}{\nu_1}\right) (\vec{e_1}\vec{e_2})^2
\sum_f | \langle f | e^{i\vec{\chi}\vec{r}} | i \rangle |^2 \times
\delta(\Delta E_{if} - \Delta h\nu) 
\end{eqnarray} 
Summation in the expression (\ref{seq}) is performed over all possible final
states of the electron (including continuum). In order to calculate the 
crossection for each of the channels (Rayleigh, Raman, Compton
scatterings) one has to substitute appropriate wave functions in the
expression (\ref{seq}). For hydrogen atom calculations can be done
analytically. Initial state will always be ground state of hydrogen with
wave function $\psi_i=\frac{1}{\sqrt{r^3_b\pi}}e^{-r/r_b}$. For Rayleigh
scattering $\psi_f\equiv\psi_i$. For Raman or Compton scatterings $\psi_f$
corresponds to one of the excited discrete or continuum states of 
hydrogen respectively.

One can show that for any given angle the total scattering crossection
(integrated over all possible energies of the photon in the final state)
does not depend on the energy of the initial photon (if factor $\frac{\nu_2}{\nu_1}$can be replaced with unity):
\begin{eqnarray}  \label{seqo}       
\frac{d\sigma}{d\Omega} = \int \frac{d\sigma}{d\Omega d h
\nu} dh\nu \approx r_e^2 (\vec{e_1}\vec{e_2})^2
\sum_f | \langle f | e^{i\vec{\chi}\vec{r}} | i \rangle |^2 = r_e^2
(\vec{e_1}\vec{e_2})^2  \sum_f \langle i |
e^{i\vec{\chi}\vec{r}} | f \rangle 
\langle f | e^{-i\vec{\chi}\vec{r}} | i \rangle
\end{eqnarray} 
Using the know relation  $\sum_f | f \rangle \langle f | = 1$ (e.g. \cite{schiff})
one can show that $\sum_f | \langle f | e^{i\vec{\chi}\vec{r}} | i \rangle
|^2 = \langle i | 1 | i \rangle = 1$. Thus in the considered range of
photon energies ($E_b\ll h\nu_1\ll m_e c^2$), angular distribution of scattered
photons will always follow the expression $\frac{d\sigma}{d\Omega} = r_e^2
(\vec{e_1}\vec{e_2})^2$ (i.e. the same as for scattering by free electron in
Thomson limit). The presence of different final states of electron 
affects only the spectrum of scattered radiation. Averaging over the 
polarization directions leads to the appearance of the factor
$(1+cos^2\theta)$ -- i.e. the Rayleigh scattering diagram.

\subsection{Rayleigh scattering}
\begin{figure}
\hbox{
\centerline{
\epsfxsize=7cm
\epsffile{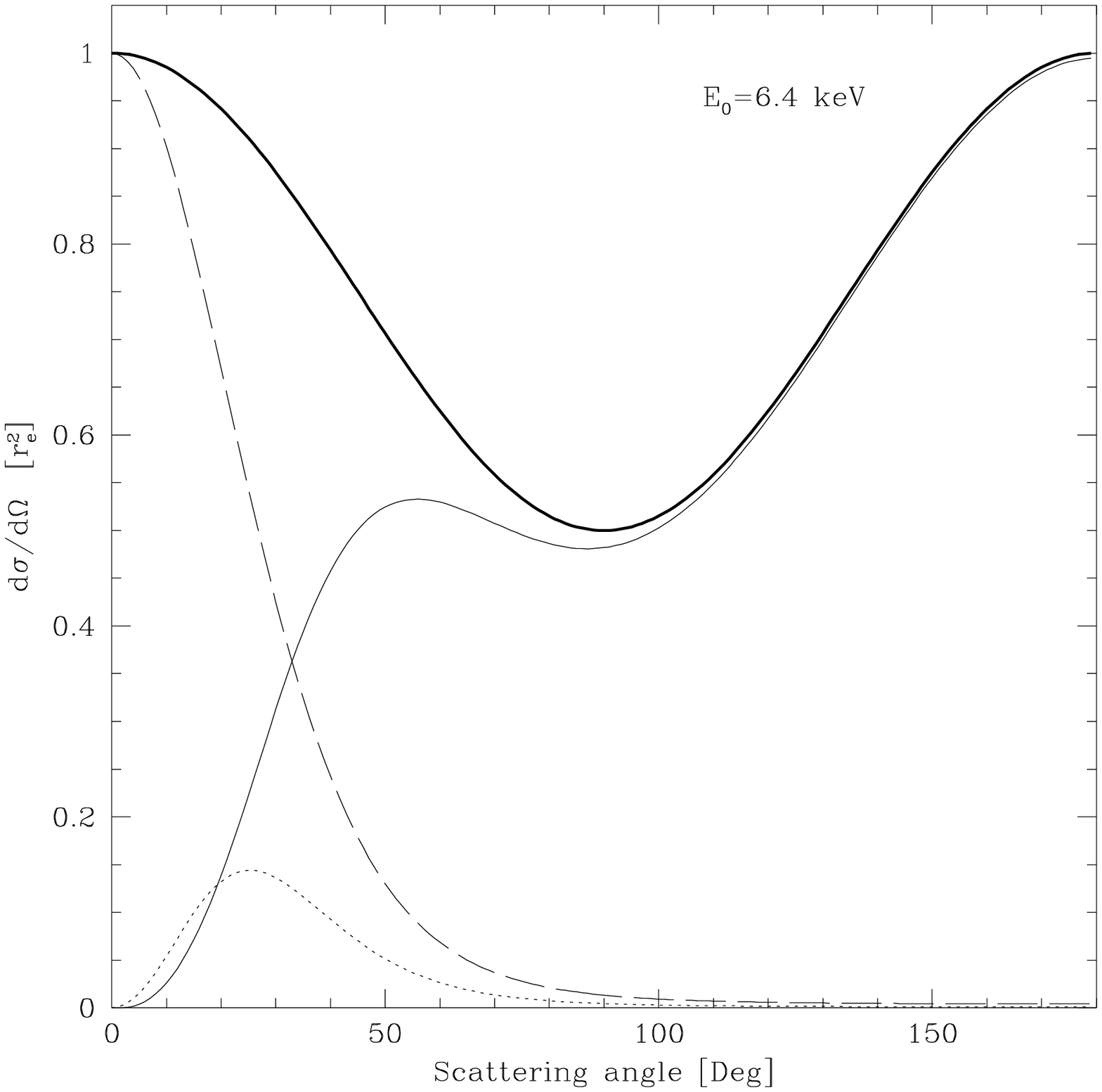}
\epsfxsize=7cm
\epsffile{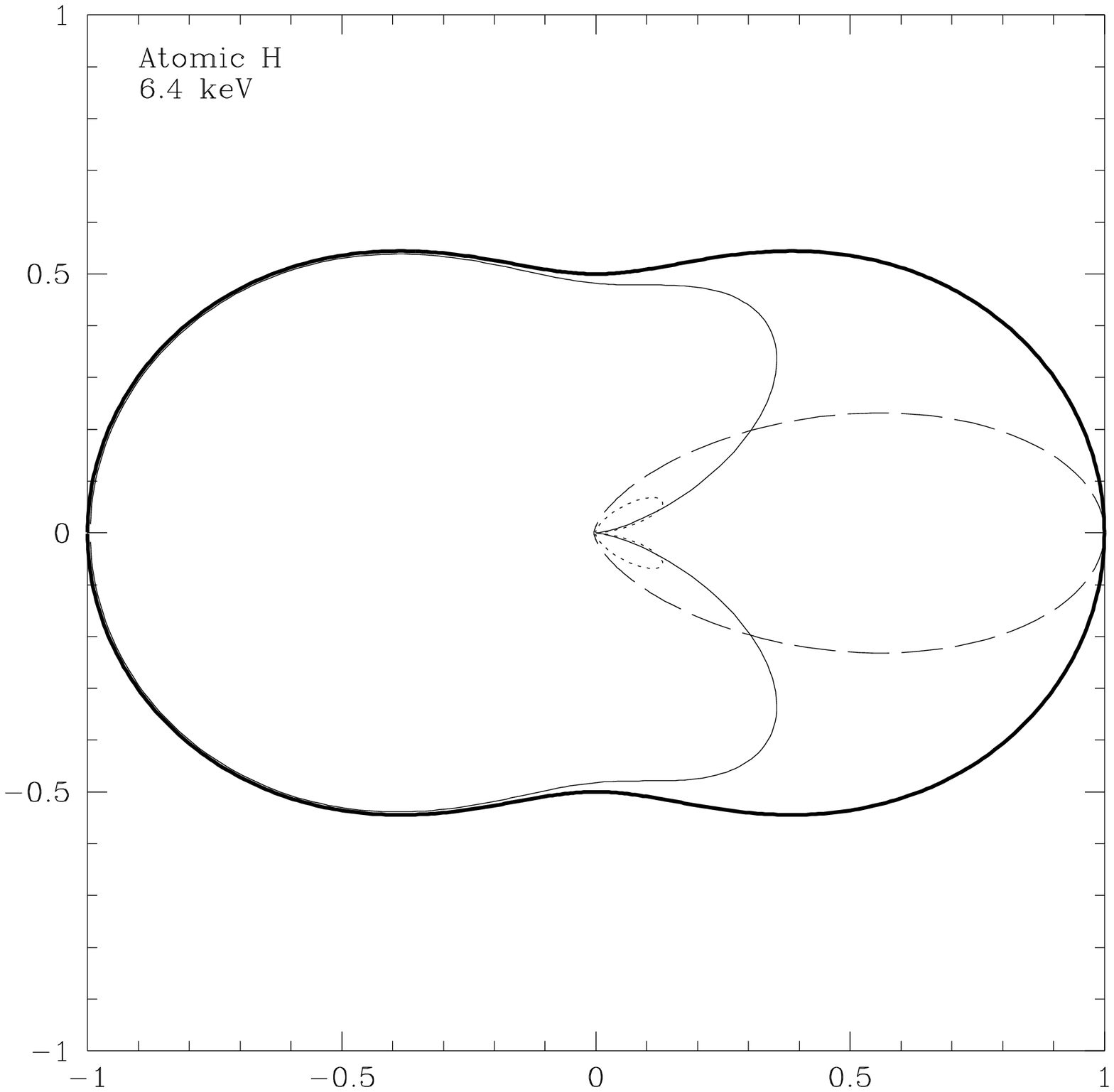}
}
}
\caption{
Differential crossection of the 6.4 keV photons scattering by atomic
hydrogen as the function of angle. Thin lines show the contributions of
Rayleigh (dashed), Raman (dotted) and Compton (solid) scatterings. Thick
line shows the crossection for scattering by a free electrons in the Thomson
limit. On the right panel the same values are shown in units $r^2_e$.
The initial direction of the photon coincides with X-axis.
}
\label{indh}
\end{figure}

{\it a) Hydrogen atom} \\
For Rayleigh scattering the final state of the electron coincides with it's
initial (ground) state. The direction of the photon propagation changes, but
the frequency remains the same. Motion of the heavy atom as a whole
compensates for 
the change of photon momentum. If the energy of the photon is greater
than the binding energy of the electron, but the wavelength is much larger than the
characteristic size of the atom then the differential crossection of Rayleigh
scattering coincides with that for free electron in the Thomson limit:
\begin{eqnarray}        \label{ray0}
\frac{d\sigma}{d\Omega}=\left(\frac{d\sigma}{d\Omega}\right)_{Th}
\end{eqnarray}
For the photons with energies of the order of 1--10 keV the wavelength is
comparable with the size of the atom and differential crossection of Rayleigh
scattering as follows (e.g. \cite{ep70} with reference to results
of \cite{sh34}):
\begin{eqnarray}        \label{ray}
\frac{d\sigma}{
d\Omega}=\left(\frac{d\sigma}{d\Omega}\right)_{Th}\cdot[1+(\frac{1}{2}ka)^2]^{-4}
\end{eqnarray}
        From (\ref{ray}) one can see that Rayleigh scattering plays an 
important role when the change of photon momentum is less than the typical momentum
of electron bound in an atom (i.e.  $ka\lax 1$). For X-ray photons the
initial momentum is relatively high and the condition of a small change of
momentum means scattering at small angles $\theta\lax 1/k_1a$
(see Fig.\ref{indh}). At $ka\gg 1$ Rayleigh crossection declines
as $(ka)^{-8}$. The fast decline of the crossection at $ka\gg 1$ is due to
the rapidly oscillating term in the expression (\ref{ham1}).

{\it b) Molecular hydrogen and atomic helium}

        The important property of Rayleigh scattering is the possibility of
coherent scattering of the photon by the electrons, located in the small
volume (e.g. in atom) with the characteristic size $l$. In classical
electrodynamics the parameter $x=l\cdot \chi$ (i.e. typical phase shift between
waves scattered by different electrons) defines the efficiency of
scattering. Intensity of radiation  scattered by $Z$ electrons is
proportional to 
\begin{eqnarray}   \label{sz}
\left(\sum^Z_j e^{\vec{\chi}\vec{r_j}}\right)^2
\end{eqnarray}
Obviously for $x\ll 1$ crossection will be proportional to $Z^2$. In the
opposite case of $x\gg 1$ each electron scatters waves independently and the 
crossection is simply proportional to $Z$. Similar behavior remains valid in
quantum mechanics. The crossection for scattering by multi-electron atom is
proportional to the expression
\begin{eqnarray}   \label{szm}
\sum_f |\langle f | \sum^Z_j e^{-i\vec{\chi}\vec{r_j}} | i \rangle|^2 =
\sum_f \langle i | \sum^Z_m e^{+i\vec{\chi}\vec{r_m}} | f \rangle \langle f
| \sum^Z_n e^{-i\vec{\chi}\vec{r_n}} | i \rangle = \sum^Z_m \sum^Z_n |\langle i | e^{-i\vec{\chi}(\vec{r_n}-\vec{r_m})} | i \rangle  
\end{eqnarray}
For $x\ll 1$ factor $e^{-i\vec{\chi}(\vec{r_n}-\vec{r_m})}\approx1$ at the
characteristic scale of atom and the crossection is proportional to $Z^2$. For
larger values of $x$ the crossection is proportional to $Z$ (i.e. to the number of
electrons in the atom).

In astrophysical conditions coherent scattering  increases  the role of
elements with  $Z > 1$ in comparison with atomic hydrogen (due to the factor
$Z$ per 
electron for small angle scattering). For normal cosmic abundance the 
contribution of heavy elements is not large:
summation over
all elements increases the crossection for forward scattering by a factor $\sim$
1.4 
per hydrogen atom (a similar factor appears in the expression for
bremsstrahlung emission of cosmic plasma, which is also proportional to
$Z^2$). The most important correction (about 40\%) is due to helium.  The
increase of the Rayleigh scattering crossection may be important for the
giant molecular clouds, scattering X-ray emission of X-ray sources.

\subsection{Raman scattering}
        For Raman scattering final state of the electron corresponds to one
of the excited discrete levels. Accordingly photon changes the energy by the
excitation energy of this level. For the hydrogen atom this means that the following
values of photon energy decrement are possible: $13.6\times(1-1/n^2)$,
where $n$  the 
is principal quantum number of the excited level. The crossection (with excitation of
level $n$)  is given by \cite{sh34}:
\begin{eqnarray}        \label{ram}
\left(\frac{d\sigma}{
d\Omega}\right)_n=\left(\frac{d\sigma}{d\Omega}\right)_{Th}\frac{2^8}{3}\frac{(ka)^2}{n^3}\left[3(ka)^2+\frac{(n^2-1)}{n^2}\right]\frac{[(n-1)^2/n^2+(ka)^2]^{n-3}}{[(n+1)^2/n^2+(ka)^2]^{n+3}}
\end{eqnarray}
For X-ray photons contribution of Raman scattering to the total crossection
is never large. For very small angle scattering $ka\ll 1$ crossection 
$\left(\frac{d\sigma}{d\Omega}\right)_n \propto (ka)^2$ and Rayleigh
scattering dominates, while for scattering at large angles $ka\gg 1$ and
the crossection of Raman scattering decline as $(ka)^{-8}$. Raman scattering
gives maximal contibution when $ka\approx 1$, for 6.4 keV photons
this corresponds to scattering at $\sim$30 degrees. 

It is worth mentioning again that for the scattering of a monochromatic line
with energy 
$h\nu_1$ a set of monochromatic lines will appear after the Raman scattering
with energies $h\nu_2=h\nu_1-\Delta E_n, n=1,2,..$ (see
Fig.\ref{spec}). This opens the possibility to observe the energy gap of
10.2 eV 
(the energy corresponding to the $1s-2p$ transition in hydrogen) just below the
energy of the initial photons. Scattered photons can not appear in this gap
because of the energy conservation law. 

\subsection{Compton scattering}
For Compton scattering final state of the electron corresponds to the
continuum. For the scattering by a free electron at rest the energy of the
scattered photon is unambiguously related to the scattering angle
according to (\ref{recoil1}). For the scattering by a bound electron this
unambiguous relations breaks even if the atom or molecule are not moving
before scattering. This is because the electron is not ``at rest'' in the atom,
but has certain distribution over momentum. 
For 
scattering at 
large angles (where Compton process dominates) energy transfer is greater than
the binding energy of the electron. In such condition scattering process is
``fast'' compared to the motion of the electron in atom, and the whole process
can be described as scattering by free electron having ``random'' initial
momentum, corresponding to electron momentum distribution in hydrogen
atom. 
There is no violation of the momentum conservation law, since heavy nucleus
compensate the initial momentum of the electron, without affecting the energy 
conservation law.
Detailed justification of the possibility of such
treatment of the scattering process (so called ``impulse approximation'') is
discussed by \cite{ep70}. For the free electron one can write:
\begin{eqnarray}        \label{recoil2}
\Delta h\nu  = \frac{k^2}{2m}+\frac{\vec{k}\cdot\vec{p_0}}{m}
\end{eqnarray}
where $\vec{p_0}$ is the initial momentum of the electron. Note that the first term
in the relation (\ref{recoil2}) corresponds to the usual recoil, and the second
corresponds to the Doppler effect. The broader the distribution of the
electron initial momentum, the large the deviations of energy change
(second term in (\ref{recoil2})) compared to pure recoil according to
expression (\ref{recoil1}).

Thus for bound electrons momentum distribution in the atom plays the same role
as the temperature does for the free electrons. The left (low energy) wing of the line,
scattered by free electrons of plasma with temperature $\sim$ 13.6 eV, will
resemble the result of scattering by neutral hydrogen (Fig.\ref{thin}). 

For hydrogen analytical expression of the differential crossection is known
(\cite{gl57},\cite{ep70})

\begin{eqnarray}        \label{compt}
\frac{d\sigma}{d h\nu
d\Omega}=\left(\frac{d\sigma}{d\Omega}\right)_{Th}(\frac{\nu_1}{\nu_2})\langle
| 
M_{fi} |\rangle^2 \delta (E_f-E_i-\Delta h\nu)
\end{eqnarray}
\begin{eqnarray}        \label{mfi}
\langle |
M_{fi} |\rangle^2=\frac{\pi^2 8^3 a^2}{p}(1-e^{-2\pi/pa})^{-1}\times
exp[\frac{-2}{pa}arctg(\frac{2pa}{1+k^2a^2-p^2a^2})]\times \nonumber \\
~[k^4a^4+\frac{1}{3}k^2a^2(1+p^2a^2)]\times
[(k^2a^2+1-p^2a^2)^2+4p^2a^2]^{-3} \\
where ~~p^2/2m=-|E_b|+\Delta h\nu \nonumber
\end{eqnarray}
For many electron atoms one can use ``impulse approximation'' in the limit
of photon energy decrement much larger than the electron binding energy
\begin{eqnarray}        \label{ia}    
\frac{d\sigma}{d h\nu
d\Omega}=\left(\frac{d\sigma}{d\Omega}\right)_{Th}\frac{1}{(2\pi)^3}\int\delta\left(\Delta
E 
- \frac{k^2}{2m}+\frac{\vec{k}\cdot\vec{p_0}}{m}\right) P(\vec{p_0}) d^3 p_0 =
\left(\frac{d\sigma}{d\Omega}\right)_{Th} J(\vec{k}\cdot\vec{p_0})
\end{eqnarray}
here $P(\vec{p_0})$ - is the probability to find the electron with $\vec{p_0}$
in the initial state. Quantity $J(q)=J(\vec{k}\cdot\vec{p_0})$ is so called
Compton profile. One can find extensive tables of $J(q)$ for multi-electron
atoms e.g. in \cite{bri75}. 

At lower energies the role of Rayleigh and Raman scattering increases
considerably 
along with the deviations of the Compton scattering crossection from the
case of a free electron at rest. For example differential crossections and
scattered component spectrum (averaged over all angles) are shown in Fig.
\ref{sulphur} for initial photon energy 3.32 keV (this energy corresponds to
the resonance line of the hydrogen-like argon). For the spectrum averaged
over all angles the fraction of Rayleigh scattering is 36\%, about 6\% of
the scattered photons form the first Raman satellite of the line
($h\nu=h\nu_0 - 10.2$ eV). 

\subsection{Scattering by molecular hydrogen and atomic helium}
\subsubsection{Molecular hydrogen}
\begin{figure}
\hbox{
\centerline{
\epsfxsize=5cm
\epsffile{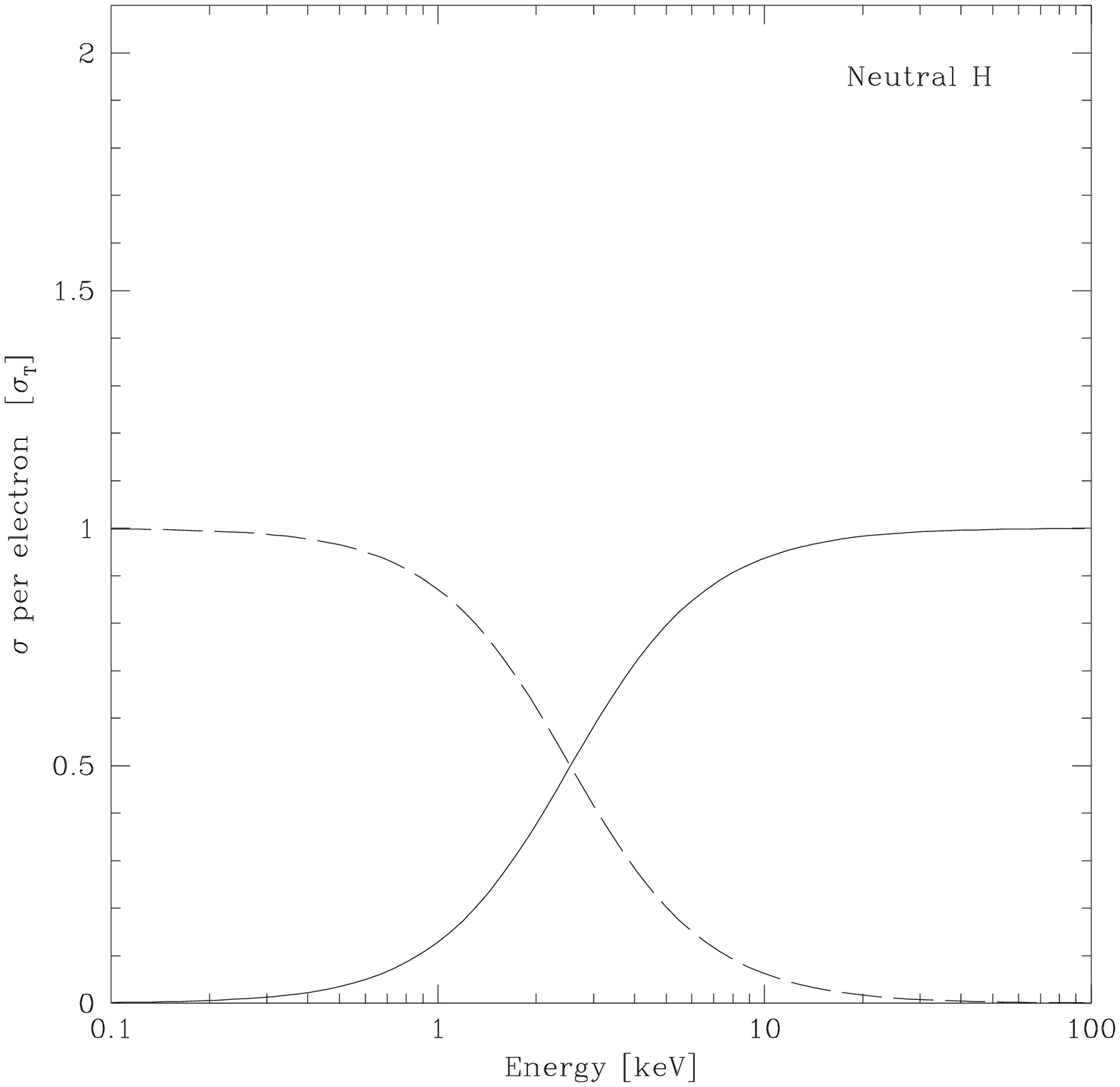}
\epsfxsize=5cm
\epsffile{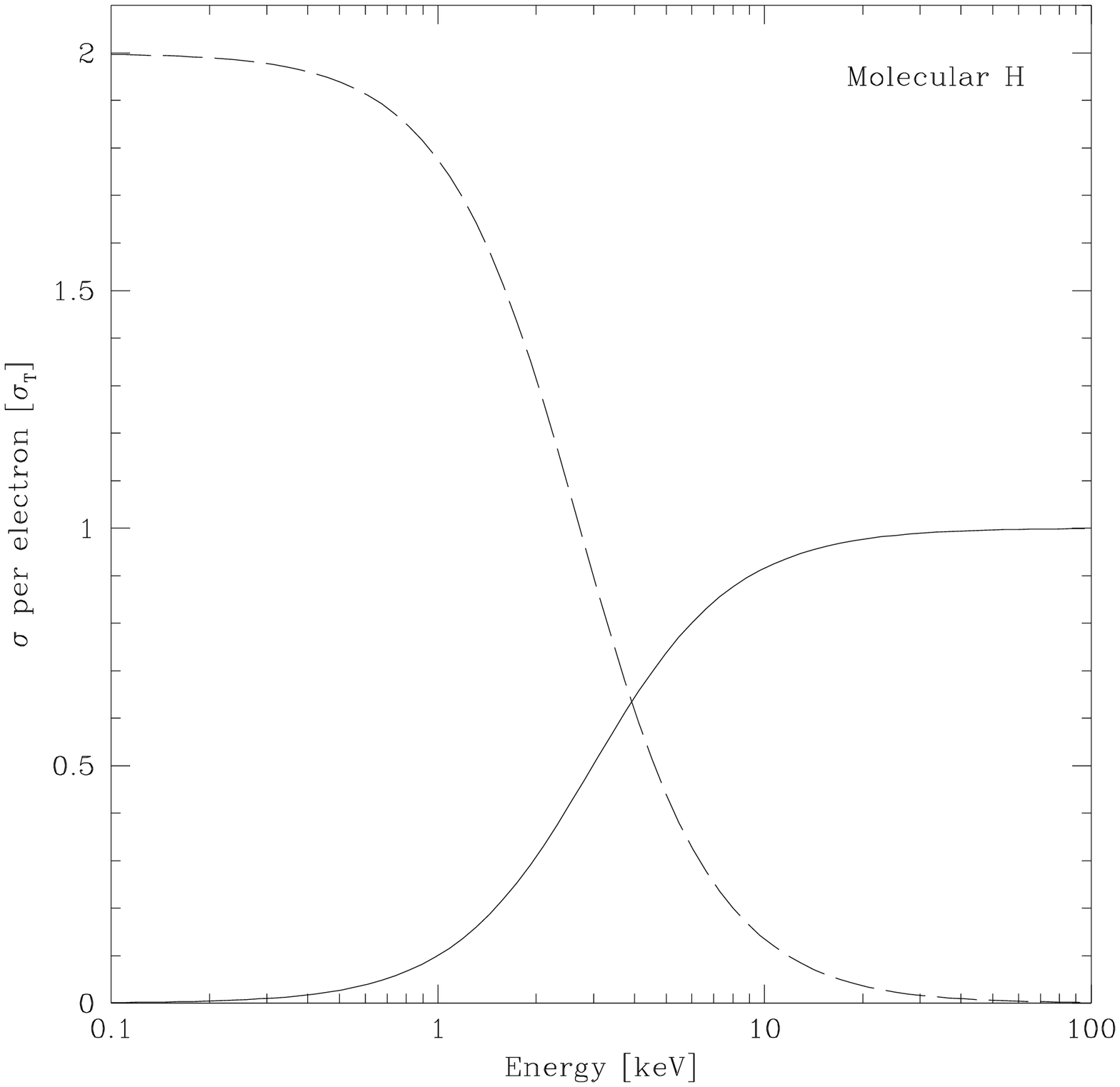}
\epsfxsize=5cm
\epsffile{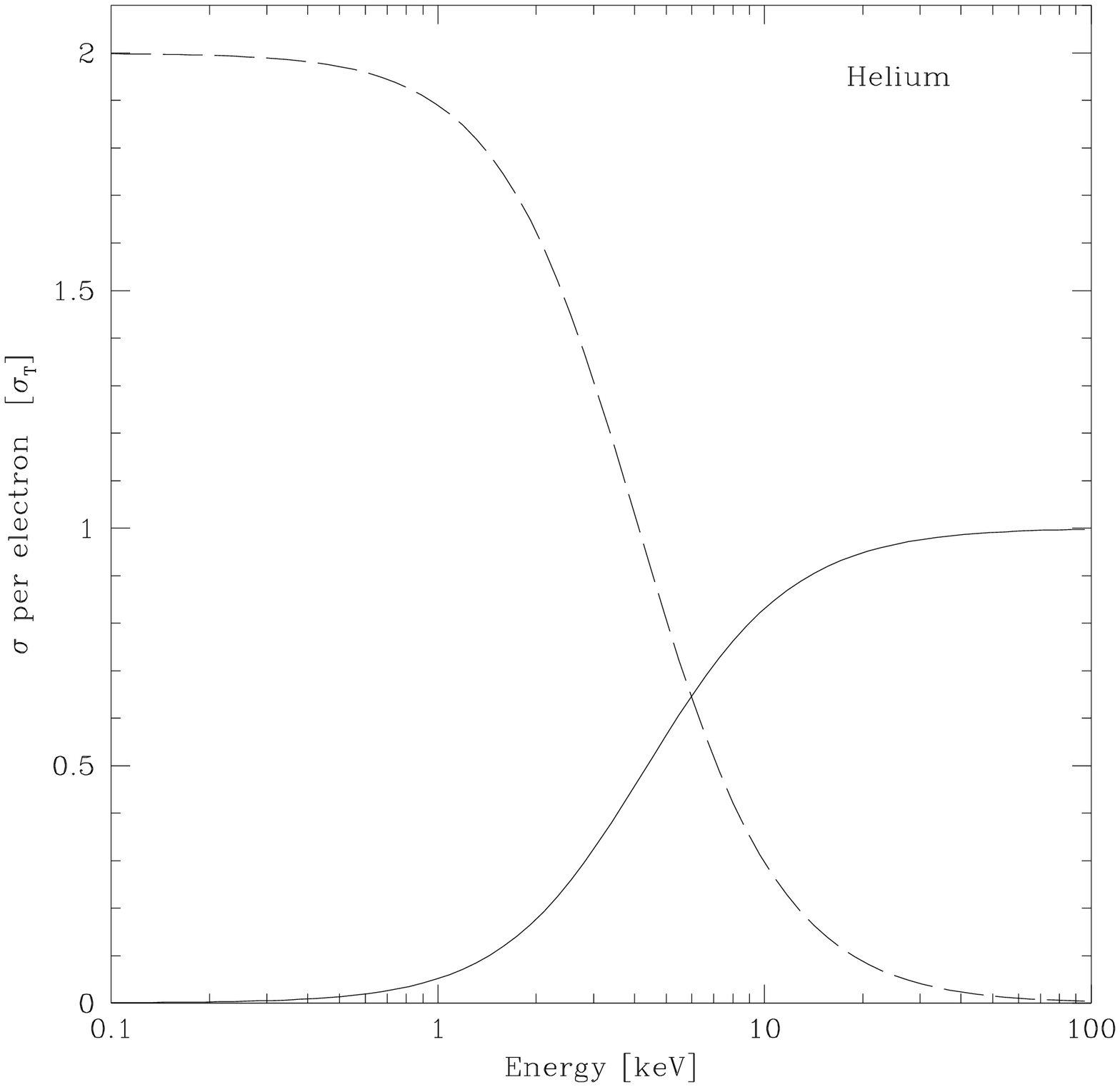}
}
}
\caption{a) The contributions of coherent (Rayleigh -- dashed line) and
incoherent (Raman and Compton -- solid line)
processes to the total crossection of photon scattering by a hydrogen atom as
the function of photon energy. b) the same as in a) but for molecular
hydrogen. c) the same as in a) for helium (using tabulated data of Hubbell
et al., 1975) 
}
\label{tot}
\end{figure}

\begin{figure} 
\vbox{
\hbox{
\centerline{
\epsfxsize=8cm
\epsffile{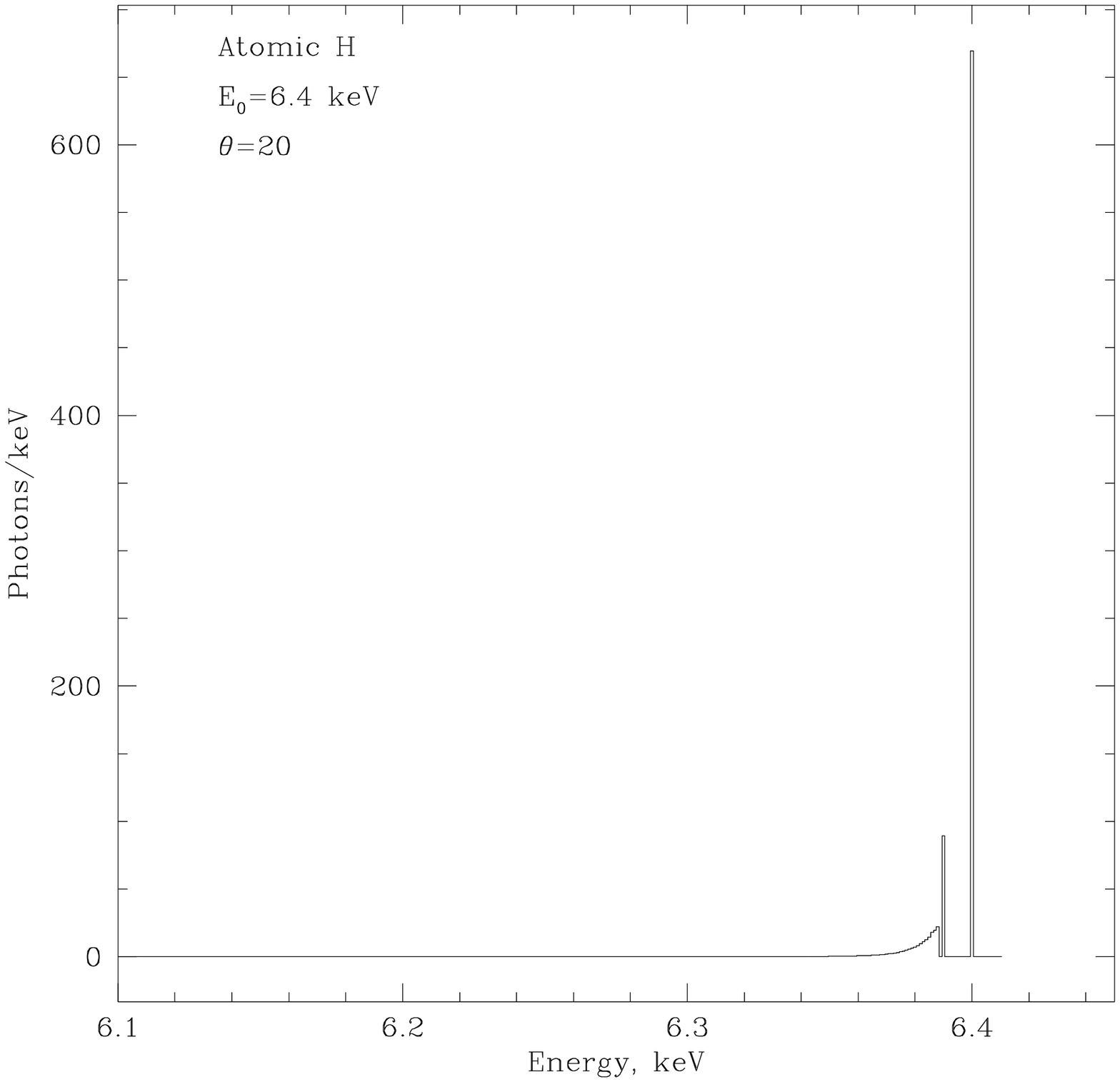}
\epsfxsize=8cm
\epsffile{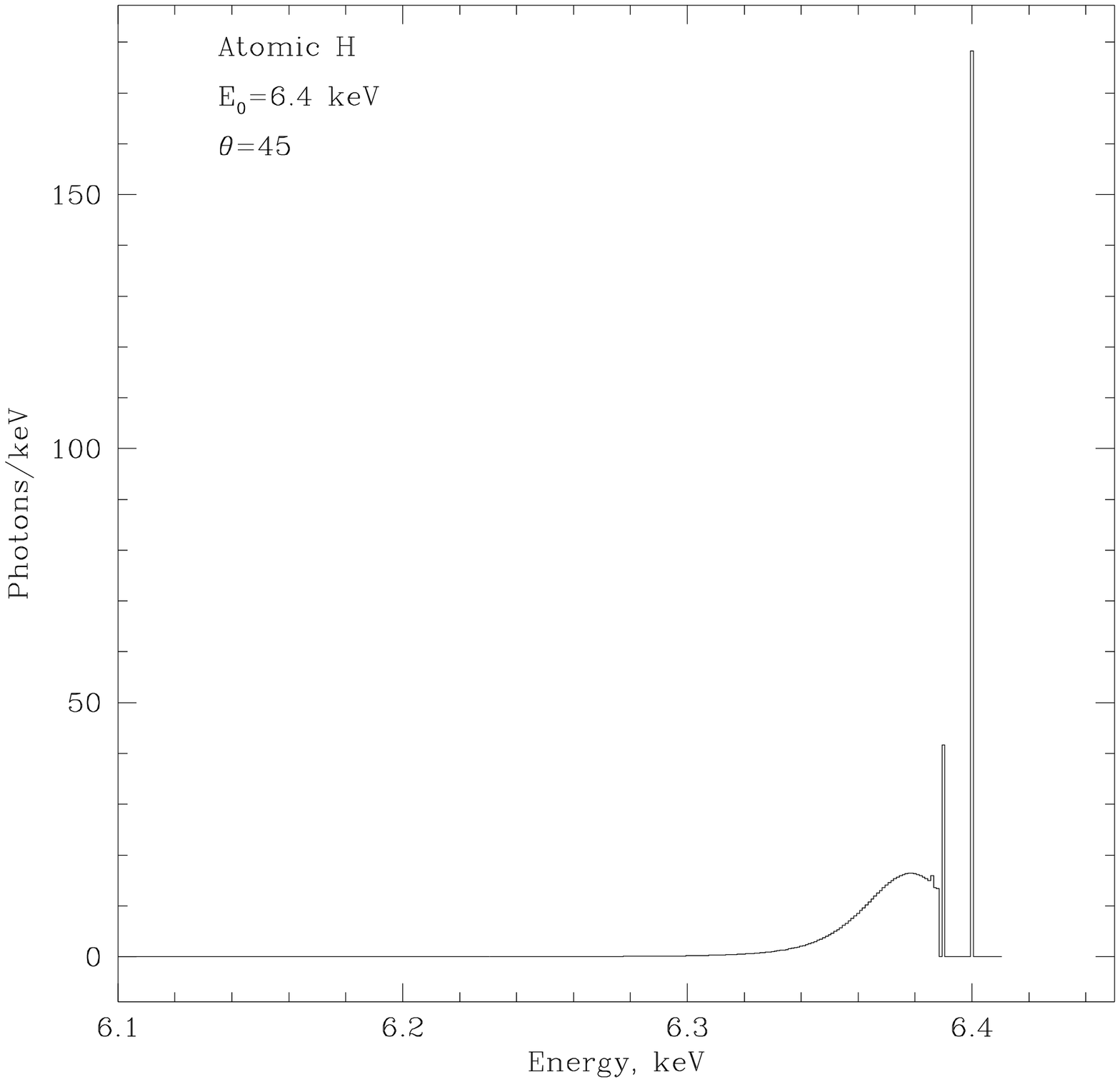}
}}
\hbox{
\centerline{
\epsfxsize=8cm
\epsffile{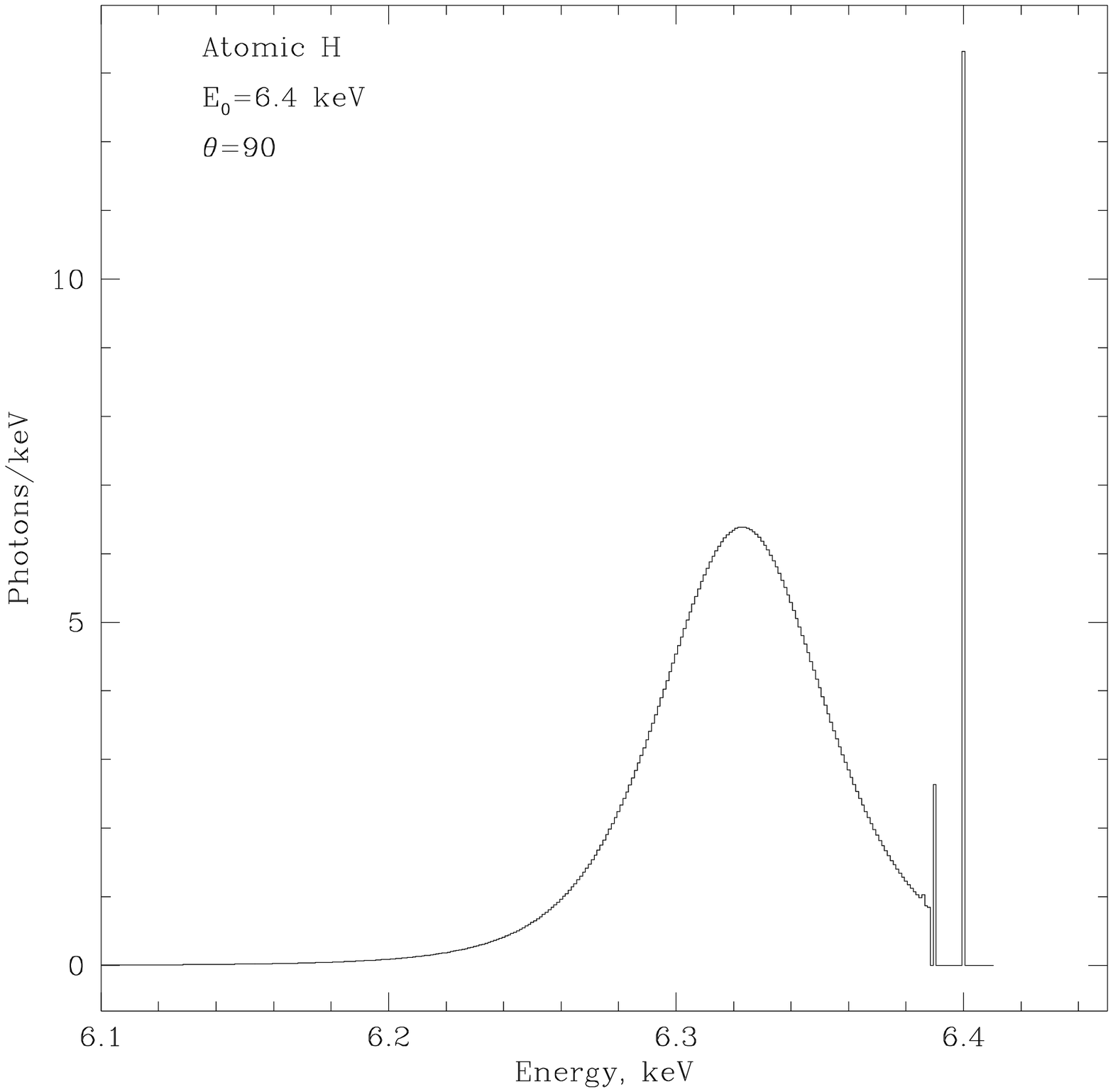}
\epsfxsize=8cm
\epsffile{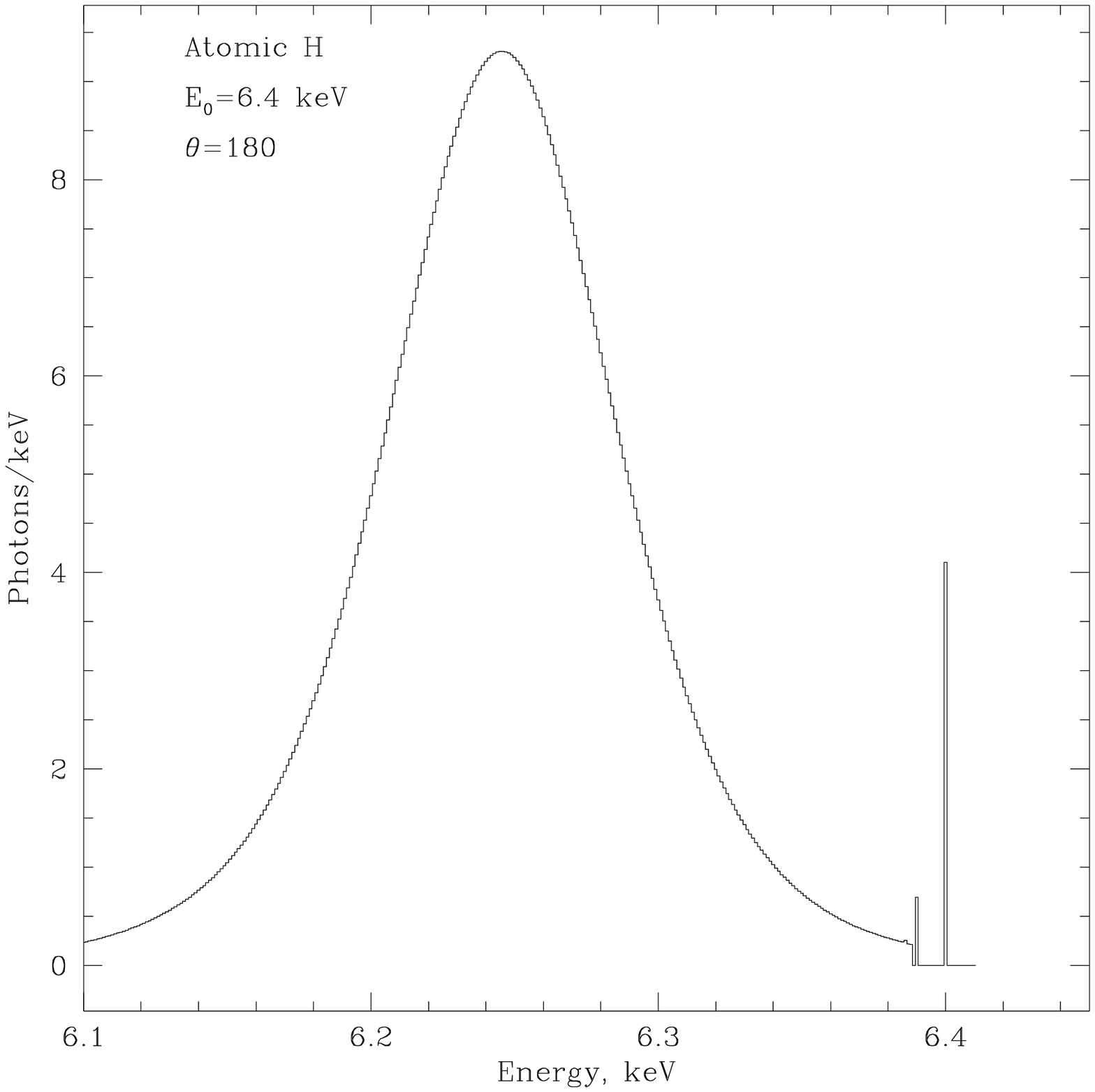}
}}
}
\caption{Spectrum of the 6.4 keV line as scattered by a hydrogen atom for
different scattering angles. The spectra are plotted with 1 eV resolution and
normalized on the same line intensity and unit interval over $\mu=cos\theta$.
}
\label{spec}
\end{figure}

The difference between scattering by molecular and atomic hydrogen
is most prominent for scattering by small angles (i.e. for Rayleigh
scattering). Firstly, coherent (Rayleigh) scattering by molecule will be
``enhanced'' by a factor of $Z$ i.e. a factor of 2 (Fig.\ref{tot}). Secondly
the 
structure of electron terms in the molecule differs somewhat from that in the
hydrogen atom. As a result the gap between the unshifted line (due to Rayleigh
scattering) and the line corresponding to the Raman scattering with excitation
of the first electron term in the molecule will be $\sim$ 11 eV compared to 10.2
eV in atomic hydrogen. 

Compton scattering by large angles will be very similar to that for atomic
hydrogen. In particular ``blurring'' of the scattered line profile due to
distribution of electron momentum in the initial state will be present. The
spectrum of 6.4 keV emission scattered by a hydrogen molecule and averaged
over all scattered angles is shown in Fig.\ref{he}. The data of \cite{e70}
on the Compton profile of the $H_2$ molecule have been used for these
calculations. Fig.\ref{he} shows that for the existing X-ray detectors
scatterings by atomic and molecular hydrogen are almost indistinguishable.  
\subsubsection{Helium}
For the scattering by a helium atom along with the increase of Rayleigh scattering
(by a factor of 2) the structure of the lines, corresponding to Raman
scattering, changes strongly. In particular the gap between ground and first
excited state in the helium atom is $\sim$20 eV. Note, that for $\sim$ 6 keV
photons 
the wavelength $\lambda\sim 2 \AA$ is comparable with the size of the atom
and the parity selection rule is not strict. Due to the stronger binding of the
electron in the helium atom, electron momentum distribution is substantially
wider than that in atomic or molecular hydrogen. As a result (see
Fig.\ref{he}) the left wing of the line will be more ``blurred''. 
A comparison of the spectra scattered by hydrogen and helium atoms
is given in Fig.\ref{he}. 
        The energy gap is twice as large as that for the hydrogen atom and
significant differences in the recoil profile allow one to hope that
studying the recoil spectrum might become a method for helium abundance
determination. Note that even after multiple scatterings photon does not
appear in this energy gap.

\section{Coefficients of transfer}
{\it a) Change of energy} \\
The difference in the differential crossections $\frac{d\sigma}{d\Omega d h\nu}$
for free and bound electrons 
should be taking into account considering the evolution of spectra in the
Fokker-Plank approximation. The Fokker-Plank equation uses the mean energy
change and the mean square of energy change of the photon after a large number
of scatterings. For free electrons at low temperature the corresponding
expressions have been given by \cite{rwm78}: the mean energy change of the
photon $A(h\nu)=-n_e\sigma_T c\left(\frac{h\nu}{mc^2}\right)(h\nu)$ and the mean
square of the energy change  $B(h\nu)=n_e\sigma_T c
\frac{7}{5}\left(\frac{h\nu}{mc^2}\right)^2(h\nu)^2$. Shown in 
Fig.\ref{avde} is the dependence of the mean energy change and the square of
energy change for scattering by free and bound electrons. The mean energy
change practically coincides with that for the free electron at rest (see
e.g. \cite{ep70}), while the mean square of energy change is larger in the case of
scattering by bound electrons. This is due to additional ``blurring'' of
the recoil profile due to distribution of the electron momentum in the initial
state. To the order of magnitude the mean square of energy change for
scattering by a bound electron is $(h\nu_B)^2= (h\nu)^2 +
h\nu <\frac{p^2}{2m}>$, where $h\nu$ is the recoil for the scattering by
free electron. Using the results of \cite{ep70} one can write the expression
for the mean  square of energy change per unit time as :
\begin{eqnarray}        \label{de2}
B(h\nu)=n_î\sigma_T c \left[\frac{7}{5}\left(\frac{h\nu}{mc^2}\right)^2(h\nu)^2
+\frac{4}{3}(h\nu)\left(\frac{h\nu}{mc^2}\right) 13.6 eV \right] 
\end{eqnarray}
The first term in expression (\ref{de2}) corresponds to usual recoil (see
\cite{rwm78}). \cite{rwm78} and \cite{ill79} considered the evolution of
spectra 
due to multiple scattering by cold electrons. In the case of neutral
hydrogen the additional diffusion term is to be taken into account. This term
dominates at $h\nu_1 \lax 2 keV$. For photon energies $h\nu_1 \gg 2 keV$
it can be neglected. Obviously this term is important only for the
case of the media, underabundant with heavy elements, when Compton
scattering dominates photoionization of the heavy elements (e.g. early
Universe at $3\lax z\lax 1500$). 

{\it b) Transport crossection} \\
        The value of transport crossection $\int (1-cos\theta) d\sigma$,
which characterizes the efficiency of changing photon momentum, is important
for consideration of spatial diffusion. For atomic hydrogen the transport
crossection is the same as that for free electrons (see above). For
molecular hydrogen, helium and heavier elements, coherent (Rayleigh)
scattering increases somewhat the transport crossection (recalculated per one
electron). For 2 keV photons the transport crossection increases by a factor
$\sim$ 1.5, in comparison with scattering by atomic hydrogen or free
electrons. With the increase of energy the contribution of the coherent
scattering to the transport crossection falls rapidly.

\section{Simple models}
        We consider below a few situations, where the account for
bound electrons can be important. Distortions of spectra are 
important only when a sharp feature is present in the spectrum. We considered
only the case of the iron $K_{\alpha}$ line, but scattering of other emission
lines (e.g. emission lines of helium and hydrogen like ions of heavy
elements illuminating a cloud of neutral gas) will lead to similar
spectral distortions. 

\subsection{Monochromatic source in  optically thin cloud}
\begin{figure}
\hbox{
\centerline{
\epsfxsize=8cm
\epsffile{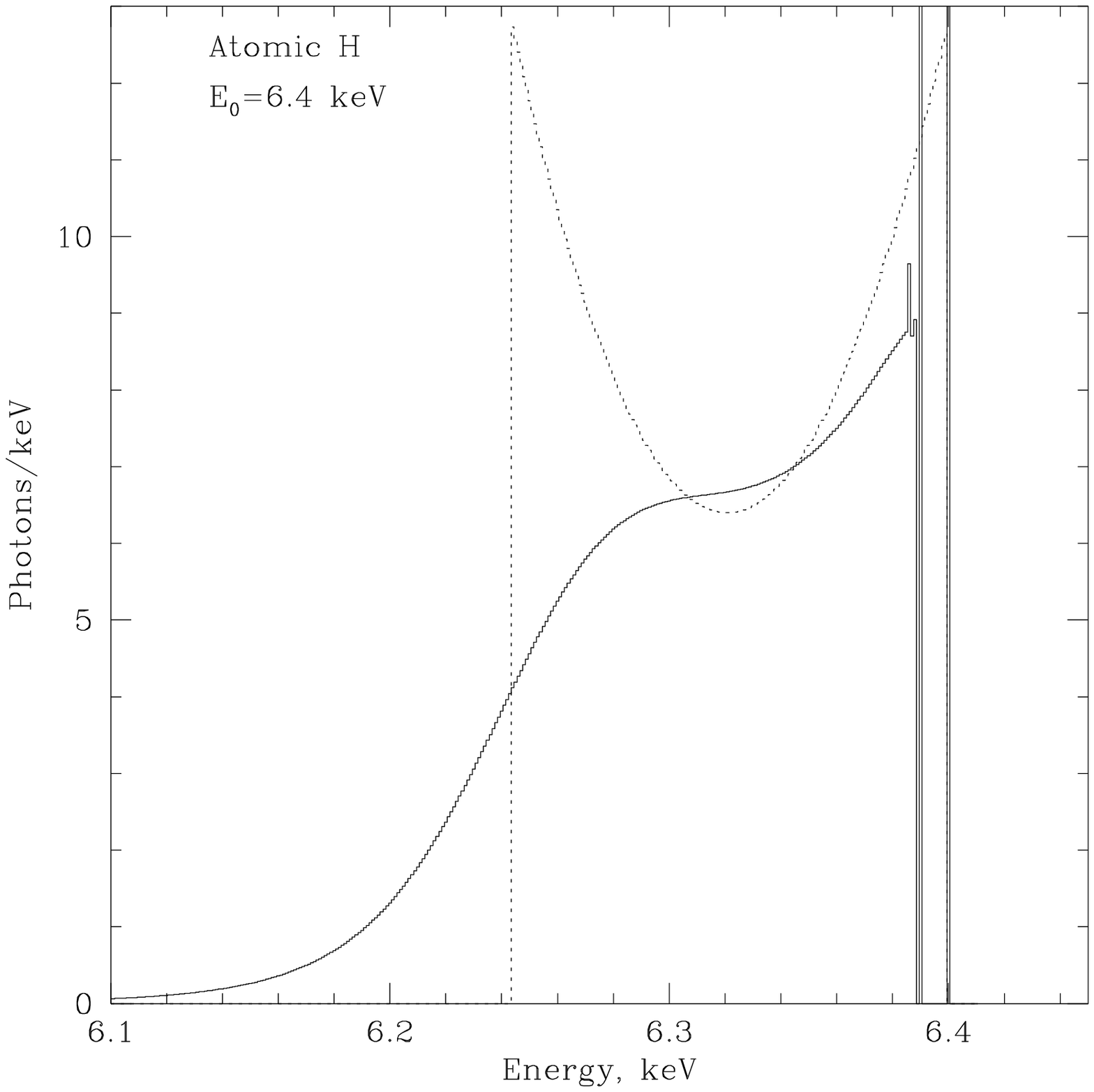}
\epsfxsize=8cm
\epsffile{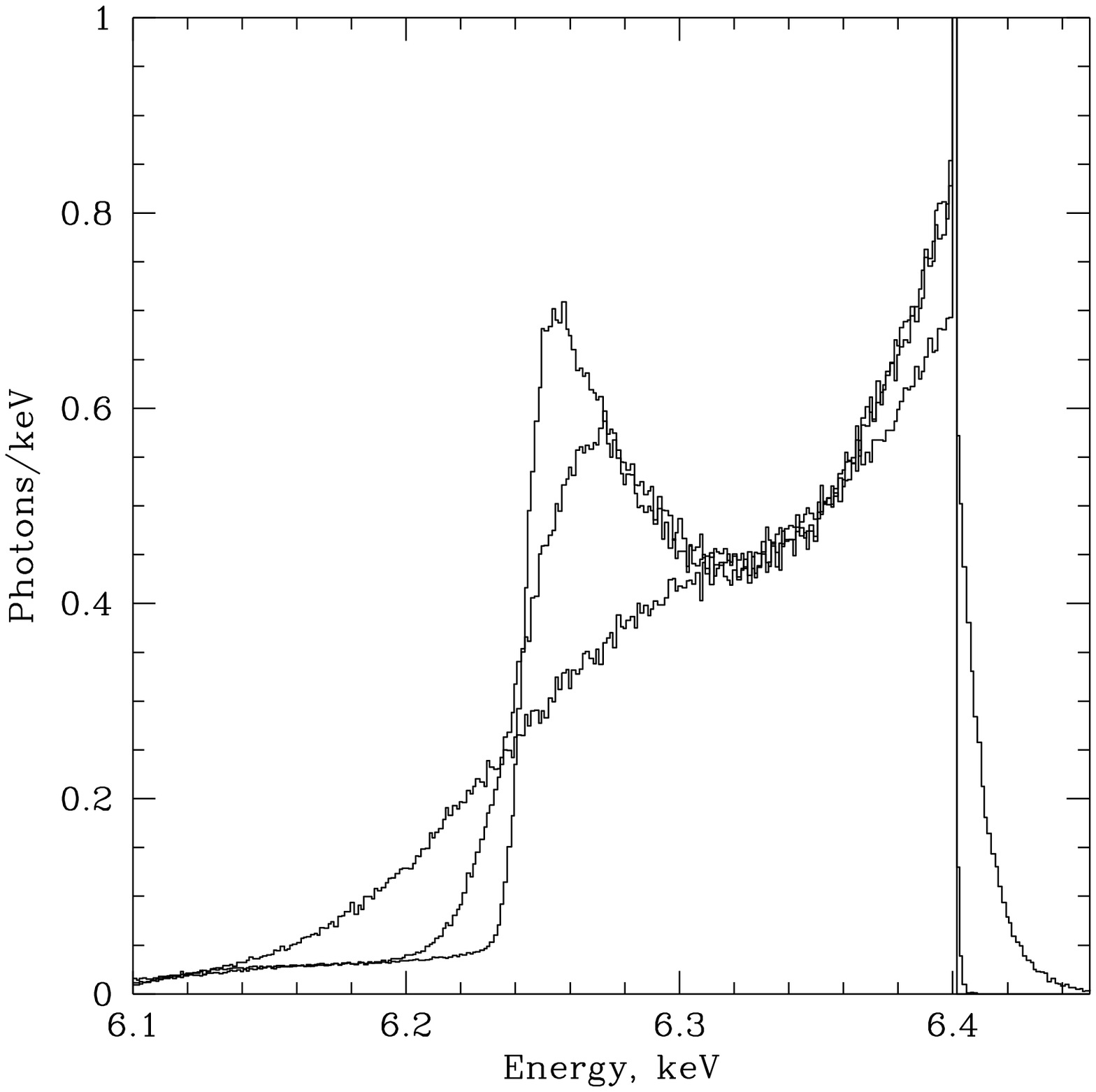}
}
}
\caption{a) Neutral hydrogen: Spectrum of the scattered 6.4 keV photons
averaged over all 
scattering angles (solid line). About 13.5 \% of photons undergo elastic
scattering, 2.3 \% of the photons change energy by 10.2 eV -- Raman
satellite of 
the line appears, corresponding to the excitation of the second level in
the hydrogen atom. The same spectrum calculated for scattering
by free cold electrons is shown by the dotted line.
b) Emission spectrum emerging from a cloud of free electrons with small
optical depth (electron temperatures are 0.1, 1 and 10 eV for three
curves respectively). Monte-Carlo simulations. 
} 
\label{thin}
\end{figure}

\begin{figure}
\hbox{
\centerline{
\epsfxsize=8cm
\epsffile{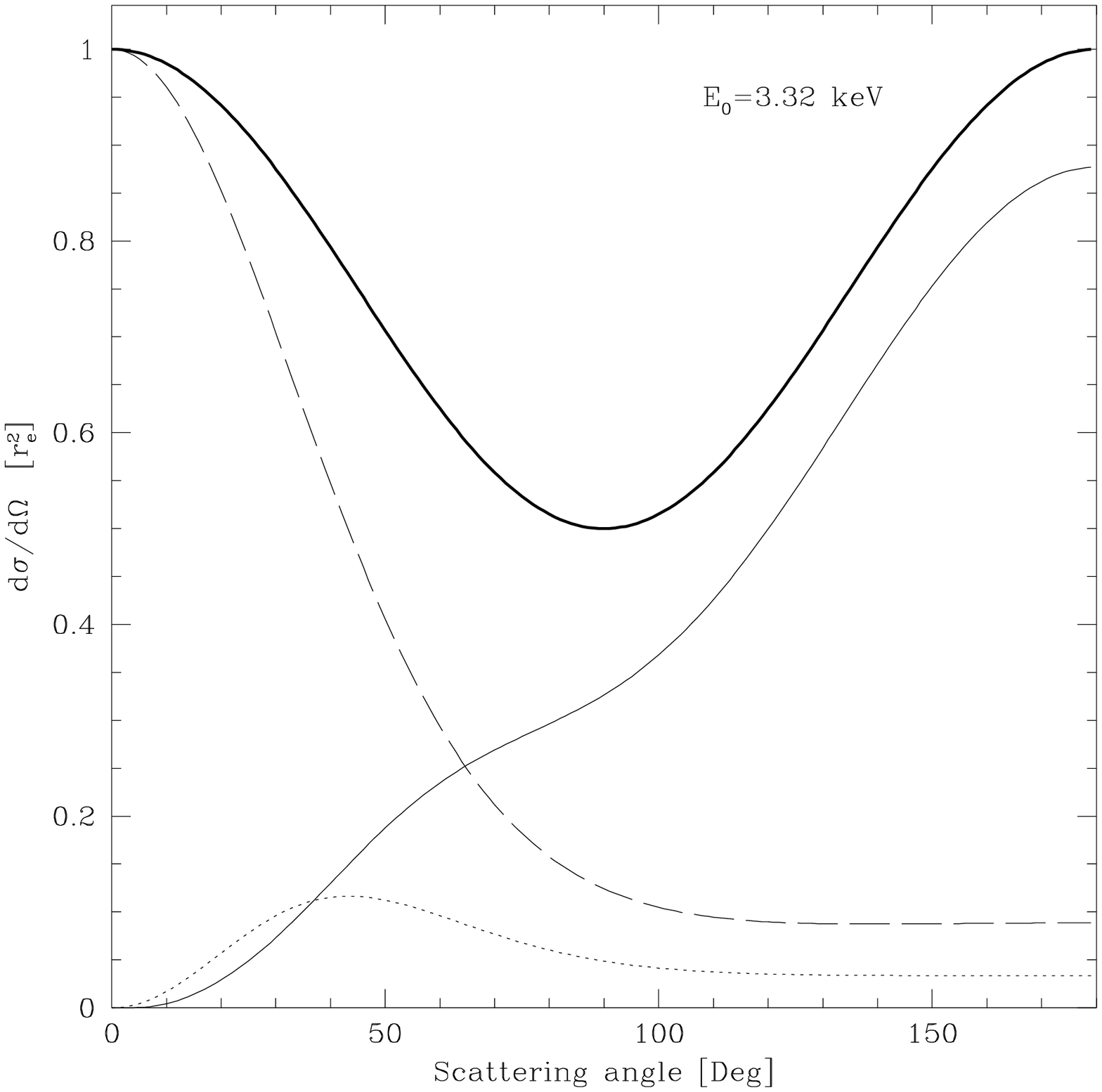}
\epsfxsize=8cm
\epsffile{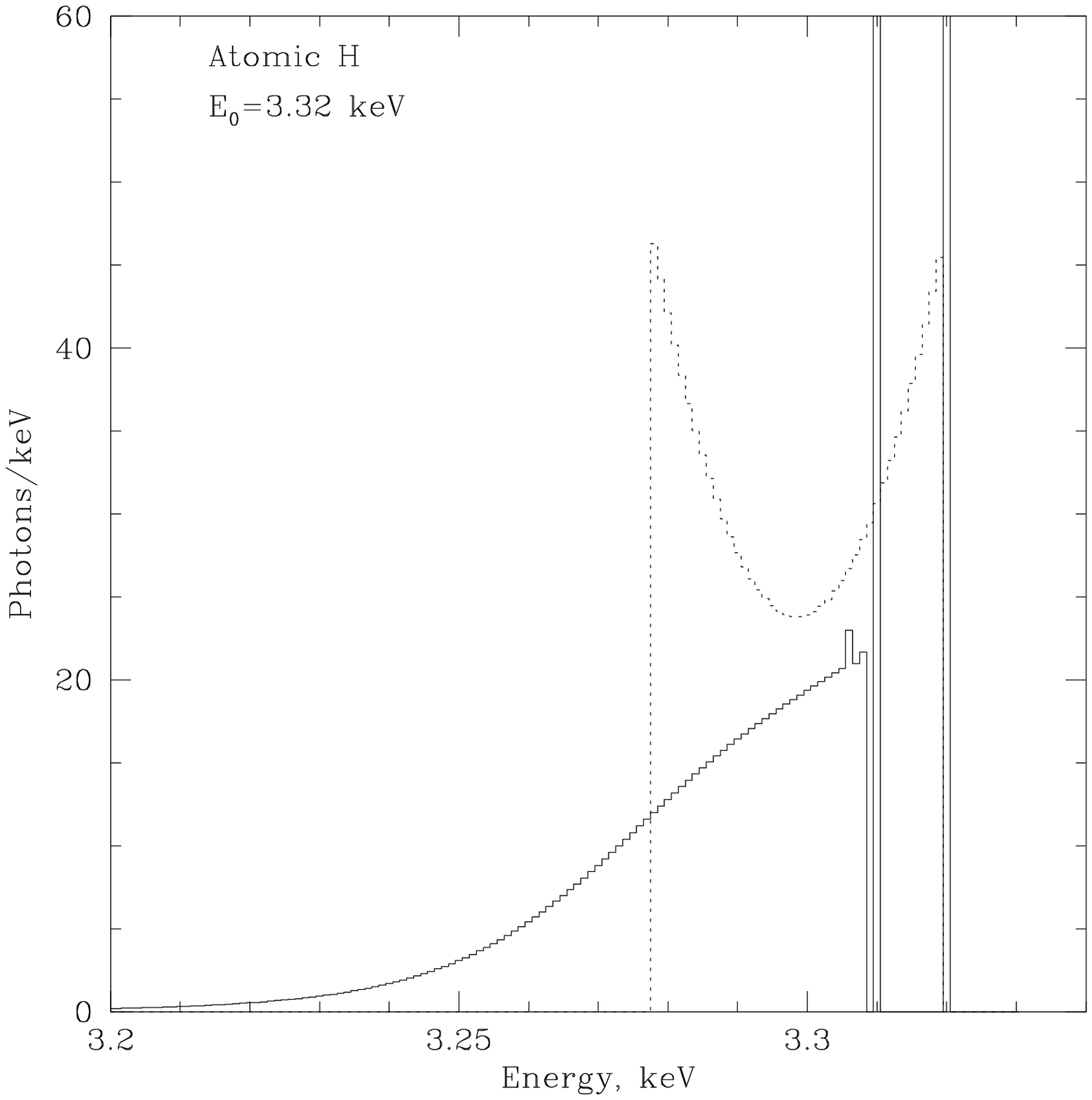}
}
}
\caption{
a) Differential crossection for scattering of the 3.32 keV photons by a neutral
hydrogen. Thin lines show contributions of Rayleigh (dashed), Raman (dotted)
and Compton (solid) scatterings. The thick solid line shows the crossection for
a free electron at rest. b) The scattered spectrum of the monochromatic 3.32
keV 
line, averaged over all directions. The dotted line shows the spectrum in the
case of scattering by free electrons.
}
\label{sulphur}
\end{figure}

\begin{figure}[t]
\hbox{
\centerline{
\epsfxsize=8cm
\epsffile{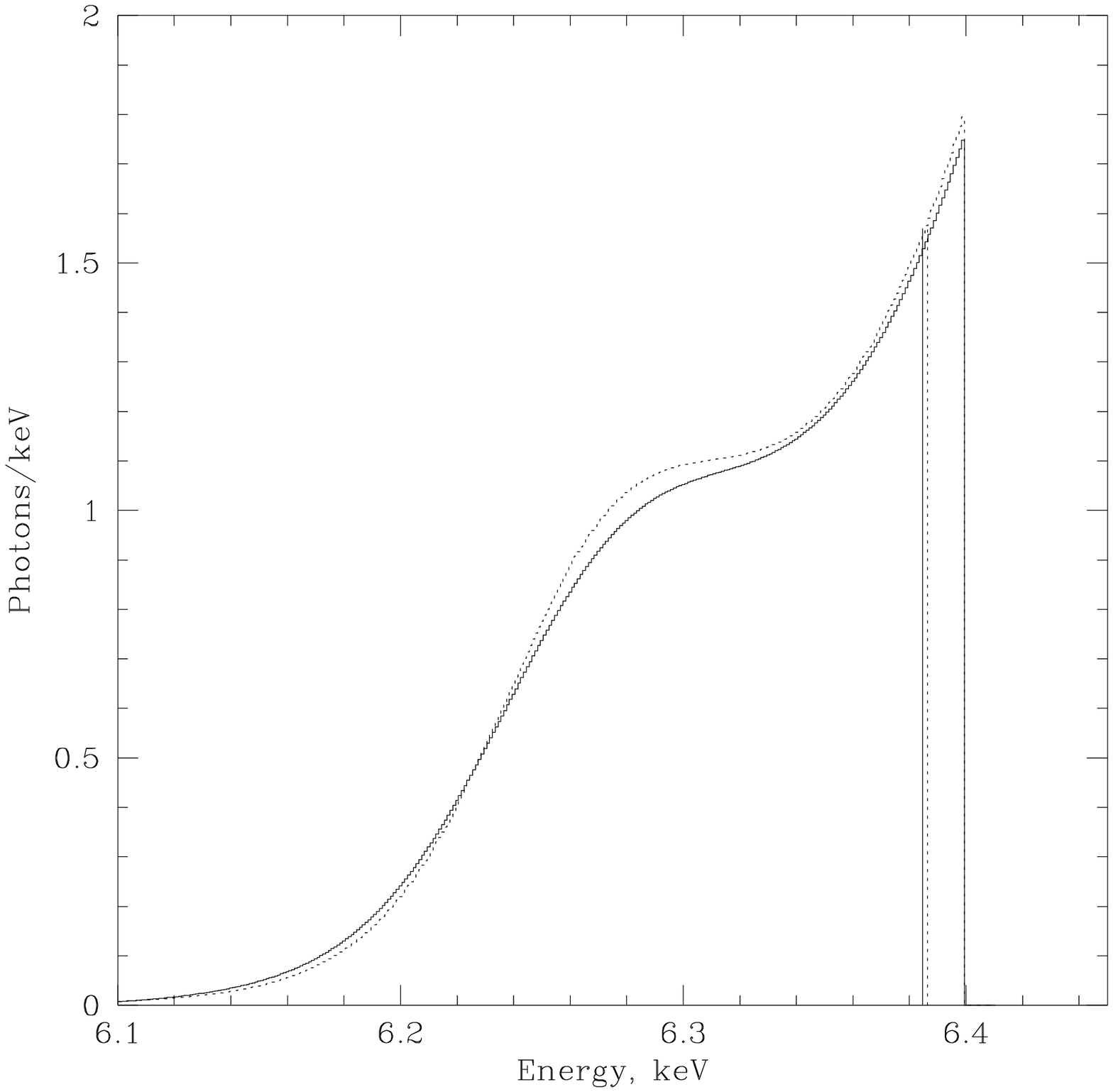}
\epsfxsize=8cm
\epsffile{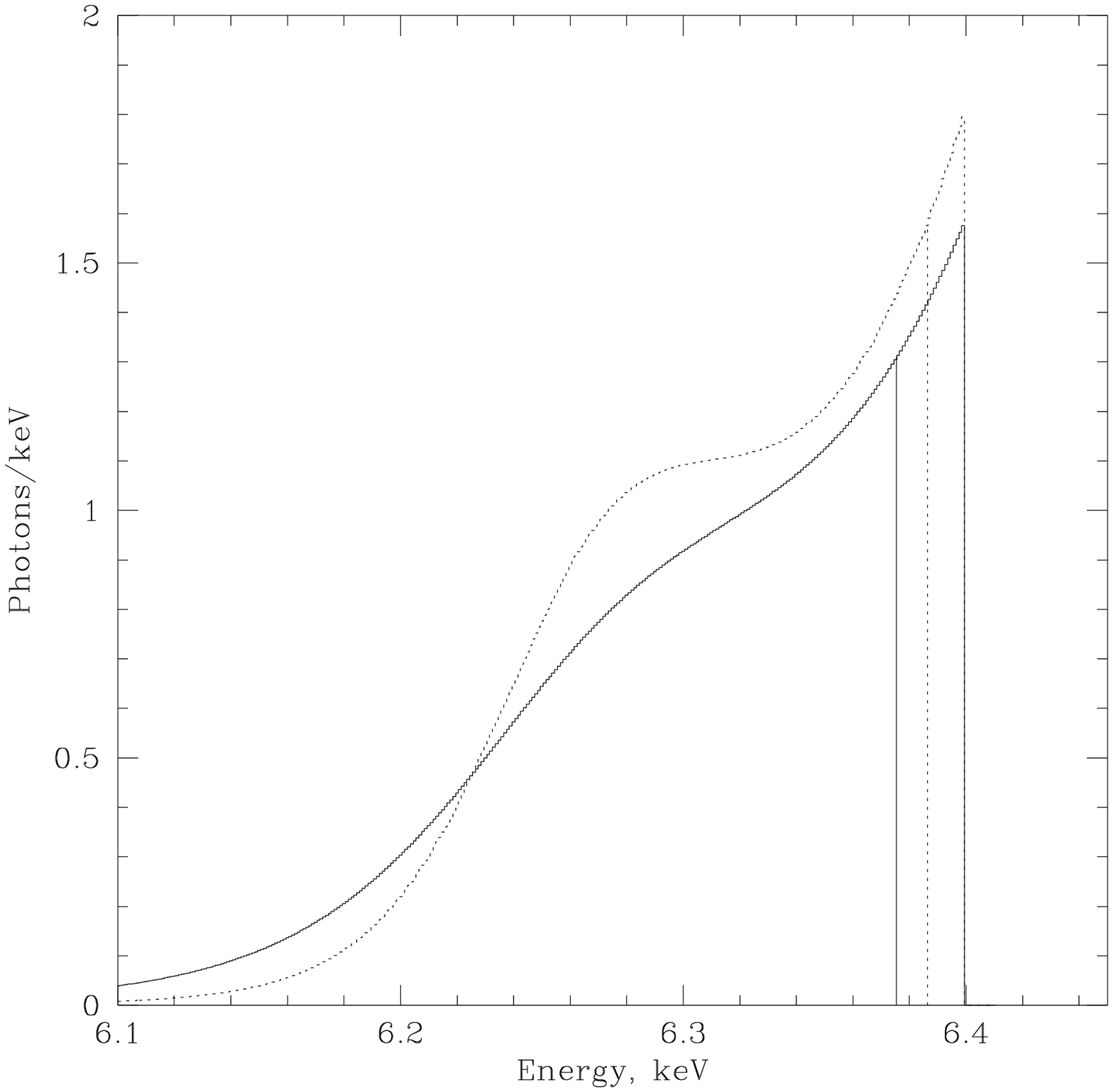}
}}
\caption{a) Spectrum of the 6.4 keV line, scattered by atomic (dotted line) and
molecular 
hydrogen (solid line) and averaged over all scattering angles. b) spectrum
of the 6.4 keV line, scattered by hydrogen (dotted) and helium atoms (solid)
and 
averaged over 
all scattering angles. The spectra have been calculated in ``impulse
approximation'' using the tables of Briggs et al., 1975. 
Vertical lines to the left of 6.4 keV separate the
regions of Compton and Raman scatterings (each line corresponds to the
energy $E=6.4-E_b$ keV, where $E_b$ is the ionization potential. ``Impulse
approximation'' is not valid to the right of these lines. 
}
\label{he}
\end{figure}

\begin{figure} 
\centerline{
\epsfxsize=8cm
\epsffile{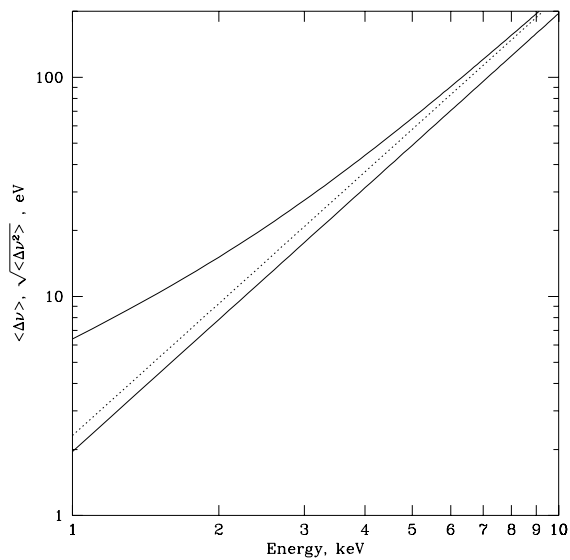}
}
\caption{Average change of energy (lower line) and square root of average
square of 
change of energy  (upper solid and dotted lines) for scattering by free
electrons and hydrogen atoms. 
}
\label{avde}
\end{figure}

All major changes in the spectrum of the radiation scattered by neutral
hydrogen can be seen clearly when considering optically thin cloud with
monochromatic source in the center. Shown in Fig.\ref{thin}a are the spectra
of a scattered 6.4 keV monochromatic line, averaged over all scattering angles ($\equiv \int \frac{d\sigma}{d\Omega d h\nu} d\Omega$)
for bound and free electrons.

        It is worth mentioning again the similarity of the distortions of the
left wing of the line for scattering by hydrogen with distortions appearing
for scattering by free electrons in warm plasma with a temperature $\sim$ 10
eV. Shown in Fig.\ref{thin}b is the spectrum of 6.4 keV photons, scattered by
free electrons with Maxwellian distribution of momenta with temperatures 0.1,
1 and 10 eV. One can see that with the increase of temperature ``blurring'' of
the left wing also increases. The right side of the line does not change much. Thus
in astrophysical conditions the line profile will be always blurred: at low
temperatures electrons are bound in atoms and the low frequency wing is
blurred due to the distribution of electron momentum in the atom, while at high
temperatures electrons are free and blurring is caused by the Maxwellian
distribution of electron momentum. Note that in the astrophysical condition
(interstellar media, stellar atmospheres, accretion disks) hydrogen is
ionized at the temperatures of the order of 1 eV. Therefore ``blurring'' of
a line may be minimal at the 
gas temperatures $T\sim$ 1--5 eV, when hydrogen is already ionized, but
the thermal velocities of electrons are still lower than the typical
velocity of a electron in a hydrogen atom.

        If the cloud is not uniform or the source is not isotropic then some
specific scattering angles will dominate. The spectral shape of the
scattered emission changes accordingly. In particular if a compact cloud is
illuminated by the emission of monochromatic photons from the distant
source, then the recoil spectrum will correspond to the scattering by the given
angle (see Fig.\ref{spec}), defined by mutual location of the source, cloud
and observer.

\subsection{Reflection of continuum spectrum by semi-infinite medium}
\begin{figure} 
\centerline{
\epsfxsize=8cm
\epsffile{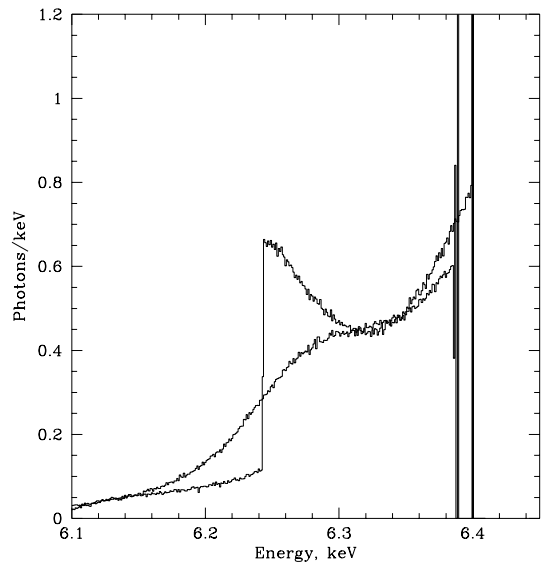}
}
\caption{Spectrum of scattered radiation in the 6.4 keV fluorescent line of
iron, 
emerging from semi-infinite medium, illuminated by the power law spectrum with
the photon index $\Gamma=2$. The spectrum was averaged over all directions. 
For comparison similar spectrum for scattering by free electrons is shown.
Monte--Carlo simulations.
}
\label{thick}
\end{figure}

        If the source of X-ray photons has continuum spectrum, then
photoabsorption in neutral medium is accompanied by the emission of
fluorescent photons in the 6.4 keV line with yield $\sim$ 0.34 per each
ionization of the electron from $K$ shell of iron (e.g. \cite{bam72}). 
        The formation of the fluorescent lines of heavy elements (mainly
$K_{\alpha}$ line of iron) due to illumination of neutral medium by the
continuum X-ray radiation have been studied in detail in a number of papers
(see 
e.g. \cite{hw77}, \cite{bas78},  \cite{bai79}, \cite{vs80}, \cite{gf91}).
Most important 
astrophysical applications of this problem 
are: Sun (X-ray continuum is emitted by the Solar flares above the surface
of the Sun), normal stars in X-ray binary systems and accretion disks with the
hot X-ray emitting region in the center. Note that angular distribution of
photons scattered by free or bound electrons is the same and it can be
described by the Rayleigh scattering diagram (see above). Thus the results of
the calculations of total equivalence  width of scattered fluorescent photons
should not alter after the account for scattering by bound electrons (note
that the contribution from Rayleigh scattering may change somewhat the
results of the 
calculations). At the same time the spectrum and angular
distribution of photons, which undergo large recoil, differs strongly. Shown
in Fig.\ref{thick} is the spectrum of the scattered 6.4 keV line, emerging from 
the semi-infinite plane of neutral matter with a cosmic
abundance of heavy elements (the approximation of photoelectric absorption
of \cite{mm83} was used) illuminated by normally incidenting
power law spectrum $F_{\nu} \sim
E^{-\Gamma} phot/s/cm^2$ with slope $\Gamma=2$. The expression (\ref{seq})
for neutral 
hydrogen was used as the differential crossection. The emergent spectrum was
averaged over all 
directions. Note again the ``gap'' $h\nu_1-10.2$~eV $< h\nu < h\nu_1 $and
the smeared left wing of 
the line at the energies $\sim$ 6.25 keV. For comparison the results of
similar calculations with differential crossection for free cold electrons
is shown as a dotted curve. This figure demonstrate the importance of the
bound electrons account for the problem of the reflection of X-rays by the
atmospheres of the Sun and late stars during strong X-ray flares. Note that
the reflection of X-rays takes place at the depth $\sim$ 1--3 $g/cm^2$, where
the degree of hydrogen ionization is low $\frac{e}{H}<10^{-3}$ (see e.g.
\cite{mal86}).

\begin{figure} 
\vbox{
\hbox{
\centerline{
\epsfxsize=8cm
\epsffile{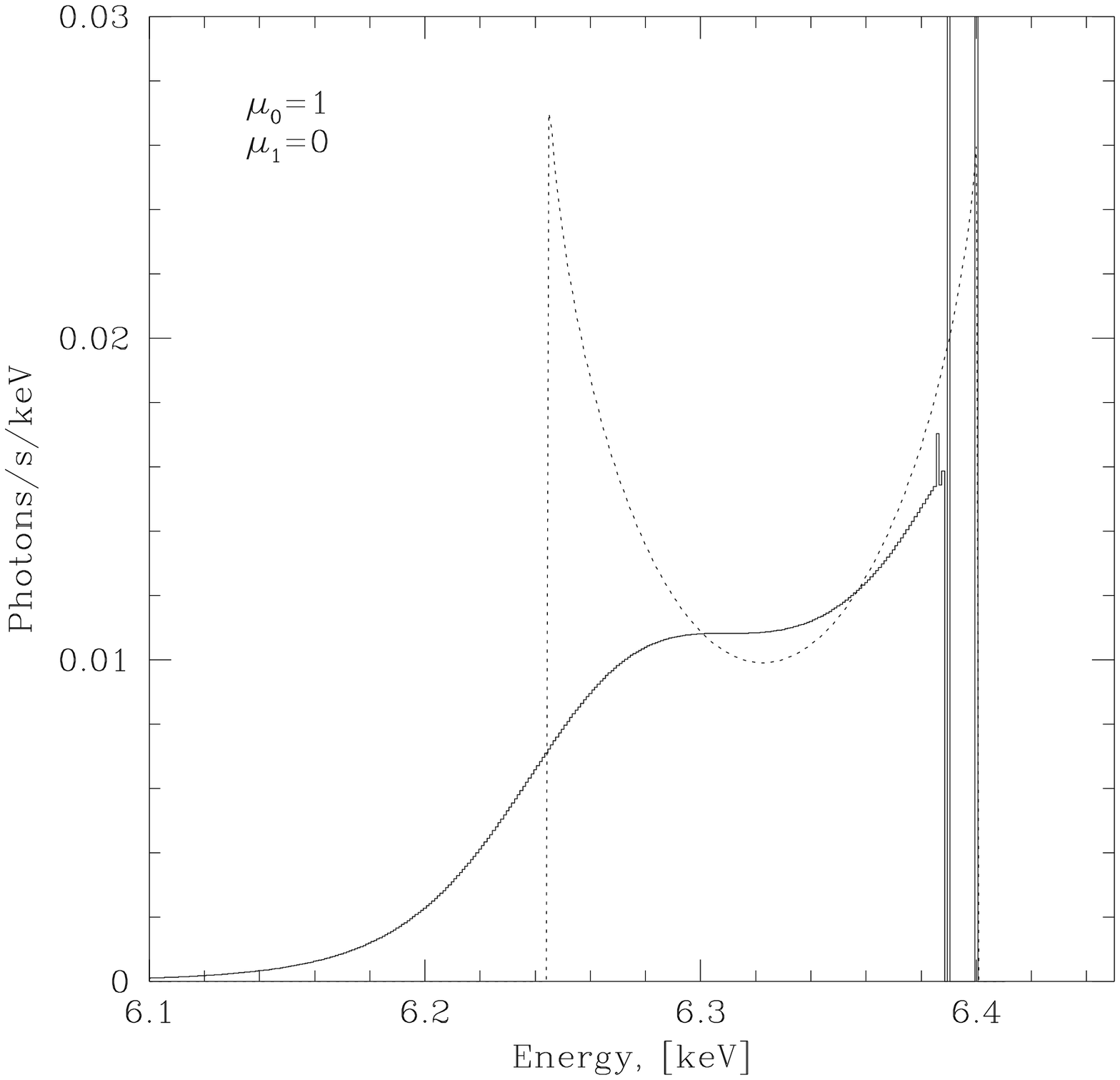}
\epsfxsize=8cm
\epsffile{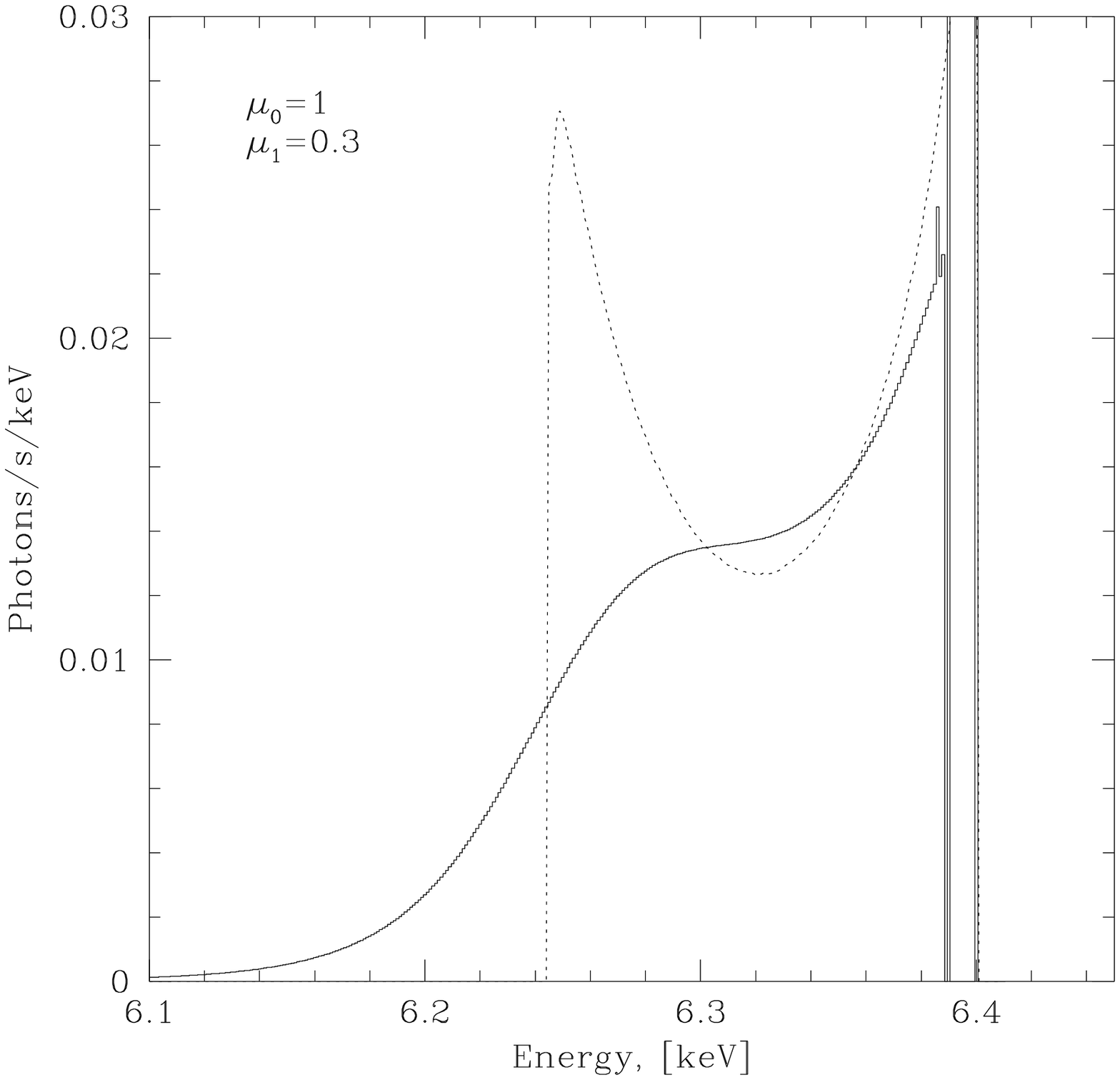}
}}
\hbox{
\centerline{
\epsfxsize=8cm
\epsffile{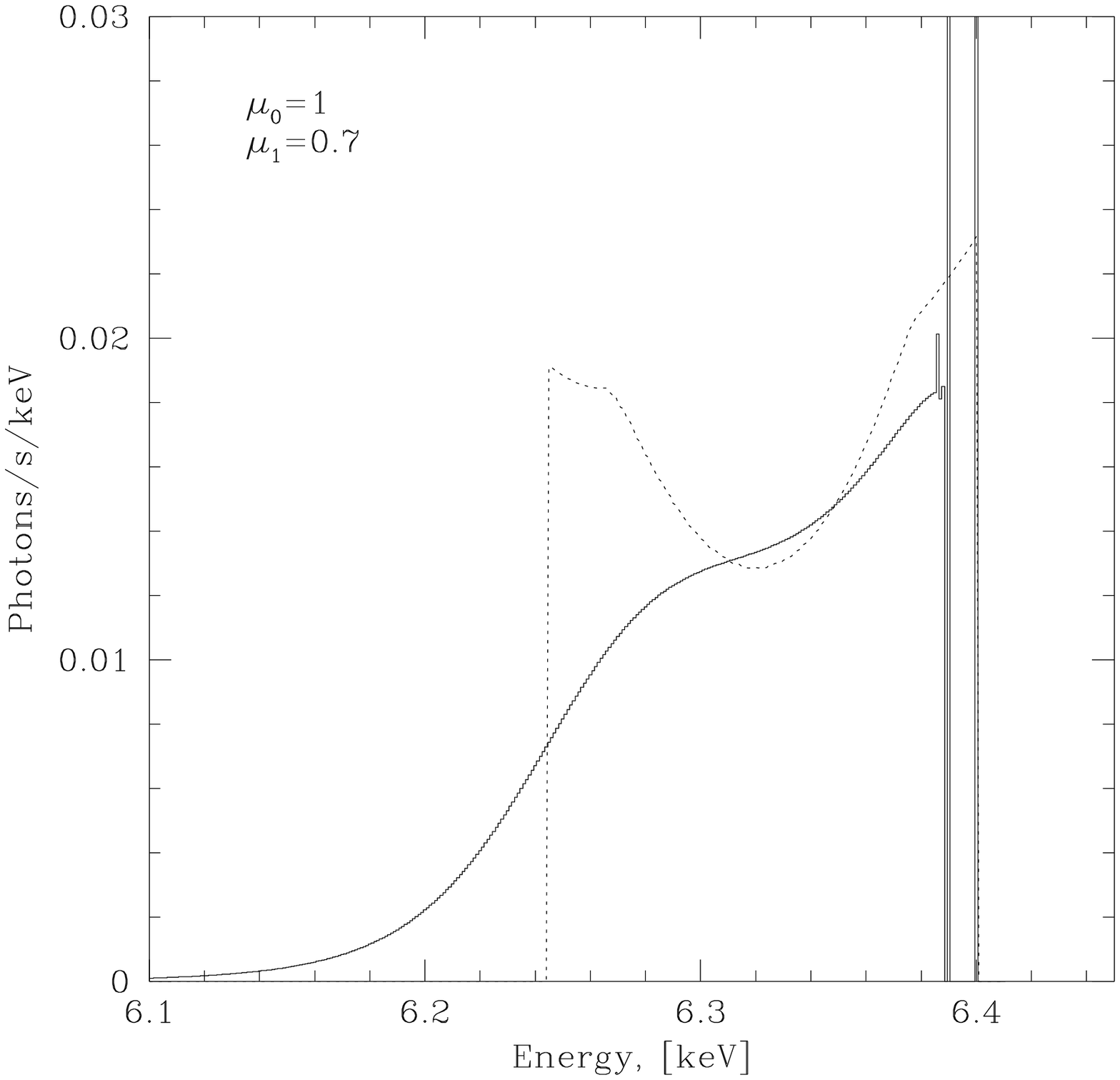}
\epsfxsize=8cm
\epsffile{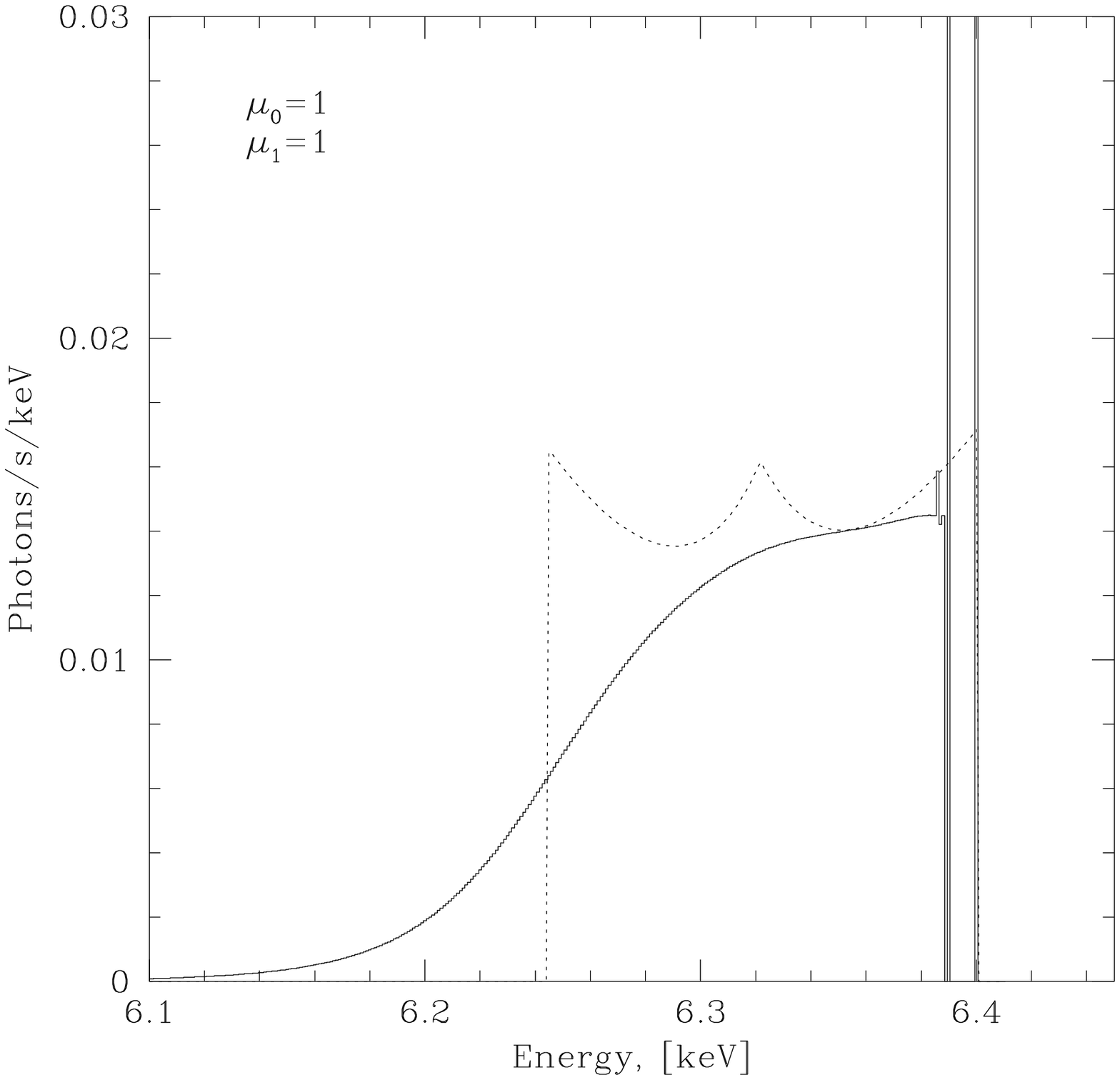}
}}
}
\caption{Spectra of single scattered radiation in the 6.4 keV fluorescent line
of iron, 
emergent from the semi-infinite media in different directions, illuminated
by continuum spectrum. $\mu_0$, is the cosine of incidence angle
for prime radiation with
respect to the direction normal to the surface ($\mu_0=1$ corresponds to the
normal incidence). $\mu_1$ is the cosine of angle between the observer and
normal to the surface ($\mu_1=1$ corresponds to the photons moving
perpendicular to the surface). Analytical approximation.
}
\label{basko}
\end{figure}

        As was expected for cosmic abundance of heavy elements, single
scattered photons gives the dominant contribution (see e.g. \cite{bas78}, ).
In the one scattering approximation the spectral shape of scattered radiation
can be calculated analytically, using the solution of the double radiation
transfer problem 
by \cite{bas78}, which assumes isotropic  scattering of continuum photons and
neglects change of their energies during scattering. Using the expression (24)
of \cite{bas78} and the differential crossection of scattering by a hydrogen atom
we obtain the spectra of single scattered $K_{\alpha}$ photons
(Fig.\ref{basko}). One can see again that the sharp backscattered peak
completely disappears.

\section{Account for the structure of iron fluorescent $K$ lines and energy
resolution of the X-ray detectors}
\begin{figure}
\epsfxsize=14cm
\epsffile{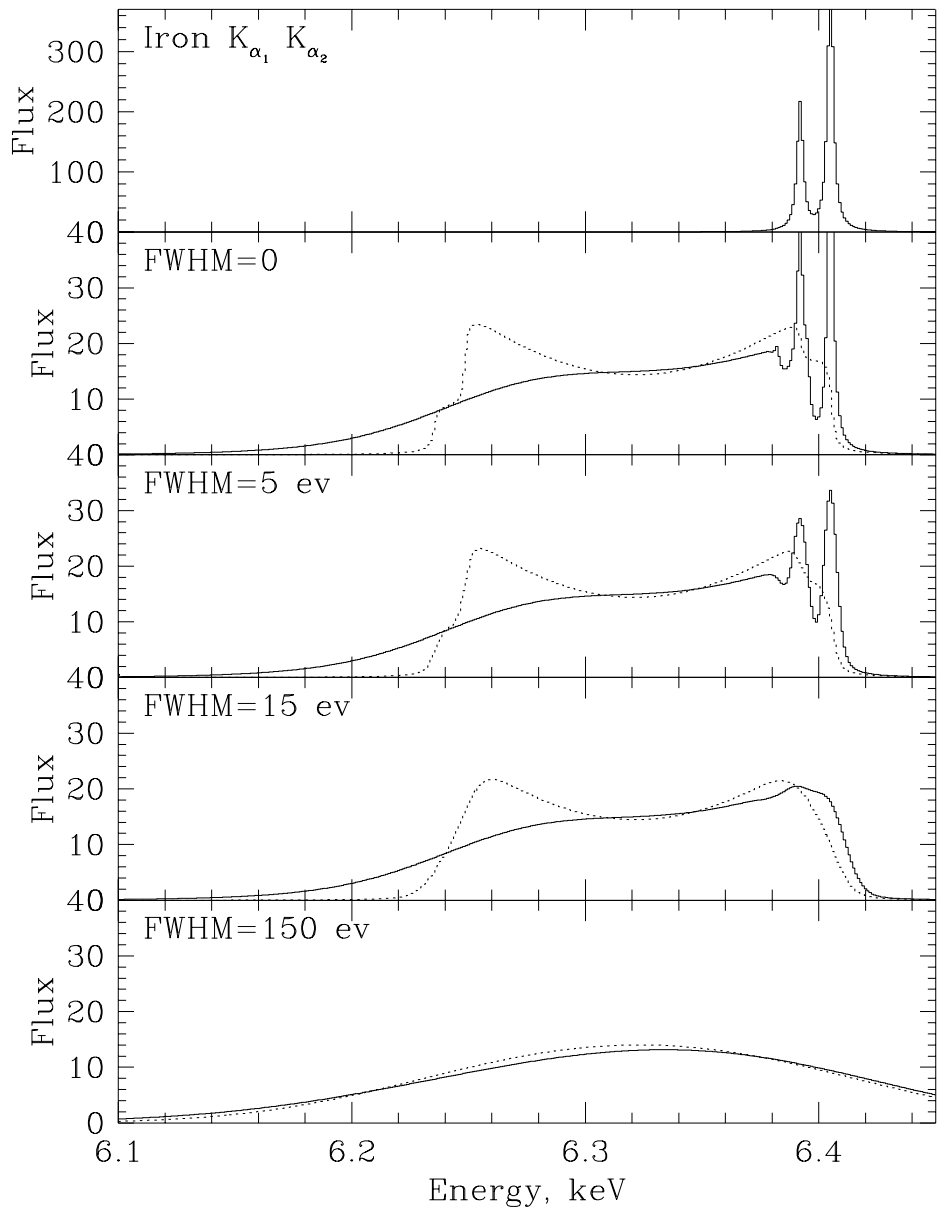}
\caption{
Structure of the iron $K_{\alpha}$ lines and predicted spectrum of scattered
radiation (averaged over all scattering angles) for different energy
resolutions of the detector. Only scattered radiation is shown. The dotted line
shows the same spectrum for scattering by free electrons. 
}  
\label{fwhm}
\end{figure}

        Previous examples deal with monochromatic lines. In order to calculate 
more realistic spectrum of scattered $K_{\alpha}$ photons one has to take
into account the intrinsic structure of iron fluorescent lines and finite
resolution of the X-ray detectors. Two lines ($K_{\alpha_1}$ and
$K_{\alpha_2}$) with energies 6.404 and 6.391 keV and relative
intensities 
2:1 contribute to the fluorescent emission of iron near 6.4 keV
(e.g. \cite{bam72}). According to interpolation of experimental data,
the intrinsic width of these lines is $\sim$ 2.65 and 3.2 keV respectively
(\cite{sl76}), although theoretical calculations predict somewhat narrower
widths $\sim$ 1.5 keV (see \cite{sl76}). Shown in Fig.\ref{fwhm} are the
spectra of the 
scattered fluorescent emission of iron averaged over all scattering angles.
Spectra have been convolved with Gaussian with FWHM=5, 15 and 150 keV in
order to demonstrate the impact of the detector resolution on the observed
spectrum. It is clear that the predicted distortions of the spectrum could be
observed by the instrument with an energy resolution worse than $\sim$ 15 eV. 

\section{Some Astrophysical applications}
Let us finally consider a few obvious astrophysical applications, where the
effects under discussion may be of interest.
\subsection{X-ray source in the molecular cloud}
	Discovery of the giant molecular cloud (\cite{bl91}) in the
direction of the bright and hard X-ray source 1E1740.7--2942 allows one to
assume that the source is surrounded by dense molecular gas. According to
the observations at millimeter wavelength the Thomson depth of the cloud can
be as high as $\tau_T\sim$ 0.2. \cite{sun91} noted that if
the source is indeed located inside the cloud, then up to 20\% of its emission
should be scattered by the gas of the cloud. The source 1E1740.7-2942 is strongly variable with the
characteristic time scale (according to GRANAT observations) of the order of
half a year. In the minimum of the light curve the flux in the hard (35-150
keV) 
X-ray band decreases by a factor of at least 5-10 (\cite{chu93}). This
variability significantly increases the chance to observe component
scattered by the molecular hydrogen. Obviously 
along with scattering, neutral gas will absorb X-rays due to photoabsorption
and emit fluorescent lines of iron and other heavy elements.

        Since optical depth of the cloud for Thomson scattering is large
enough, $\sim$ 0.2, one can hope that with the new generation of X-ray
spectrometers it will be possible to observe second order effect - recoil
effect  due to the scattering of iron $K_{\alpha}$ photons, borne in the
cloud, 
by the molecular hydrogen. This effect is proportional to the square of the
cloud optical depth: up to 20\% of the photons in the fluorescent line will be
scattered causing a decrease of the photons energies.

        Such observations might allow one to determine the position of the
source with respect to the cloud and to estimate the total mass of the hydrogen
in the cloud. The specific
recoil profile demonstrates that we are dealing with molecular or atomic
hydrogen. The fraction of the latter is not very high, since no peak of
brightness of the 21 cm line has been detected from this direction. Detailed
studies of the profile may allow one to measure the abundance of helium in
the cloud.
                
\subsection{Galactic Center region}
Another apparent example is the region of the Galactic Center as the whole.
Ginga observations have shown that central part of our Galaxy
contains a bright X-ray source, intensively emitting in the resonance
lines of helium-like iron with energies $\sim$ 6.7 keV. The ART-P telescope on
board GRANAT localized 5 compact X-ray sources within 100 pc (in projection)
from the center of our galaxy, including weak and variable hard X-ray source
within 1 arcminute from well known radio source Sgr A* (\cite{pav94}). The map
of hard 
diffuse emission in the GC region, obtained by ART-P (\cite{msp93}), agrees
well with the 
brightness distribution in the  $CO$ molecular line, which
reflects the distribution of molecular clouds. 

        \cite{smp93} noted, that such diffuse emission can be due to the
scattering of the emission from the compact sources, which have been bright
in the past, by the gas in the molecular clouds, surrounding GC.  Obviously
if Sgr A or any other compact binary source in this region was emitting
100--400 years ago at the level exceeding $10^{39}$ ergs/s, then scattering
photons will be observed today as diffuse emission. \cite{smp93} predicted
that if this hypothesis is correct, then molecular clouds should be bright
in the neutral iron 6.4 keV fluorescent line.

        This prediction has been confirmed by recent ASCA observations
(\cite{koy94}), which detected a very bright iron fluorescent line from the
direction of the large molecular complexes Sgr B, Sgr A and Sgr C. This
leads to the problem of the 
scattering of $K_{\alpha}$ photons (originating from the cloud)  in the same
molecular clouds. 
In addition ASCA results
confirmed the presence of diffuse emission in the resonance lines of
helium-like 
iron with an energy of $\sim$ 6.7 keV. Thus, molecular clouds should scatter
emission in resonance lines of strongly ionized iron, which illuminates
clouds from outside.

        As is known the data on the mass of molecular gas in the GC region
obtained from observations of the $CO$ molecule contradict  the
data on brightness of gamma-rays at the energies of a few 100 MeV originating
from interaction of cosmic rays with molecular clouds 
due to $\pi^0$ decay and bremsstrahlung of ultrarelativistic electrons. 

        The first method of mass estimation is based on the calibration of the
$CO$ flux and the amount of $H_2$ molecules using the nearby clouds and uses
observed velocity dispersion and virial theorem (e.g. \cite{sss79}). 

        The results of the second method can be agreed with those of the first
only assuming that the density of the cosmic rays in the vicinity of GC is an
order of magnitude lower than that observed in the disk (\cite{bli85}). 

        With the appearance of a new generation of X-ray telescopes with high
sensitivity and energy resolution (AXAF, Spectrum-X-Gamma, ASTRO-E, XMM)
observations of recoil profile may become an important source of information on
the amount and distribution of atomic and molecular hydrogen in this region.

\subsection{Active galactic nuclei}
Fine spectroscopy of the active galactic nuclei emission is one of the main
goals for the new generation of the X-ray telescopes. As is known, a
significant fraction of AGN spectra is characterized by the strong
absorption at low energies, which can be interpreted as due to absorption in
the molecular torus, surrounding the central source. The Thomson depth of the
torus can be of the order or above 1. Since the matter in the torus is
neutral, the profile of the lines will be disturbed due to the effects,
considered above.

Another important subject is the study of the line profiles, formed in the
accretion disks around the galactic nuclei. Doppler shift causes broadening of
the line and allows one to use the shape of the line profile for the
diagnostic of the matter motion in the disk. Scattering by neutral matter
can also contribute to the distortions of the line profile.

\subsection{Sun}
\begin{figure}
\epsfxsize=14cm
\epsffile{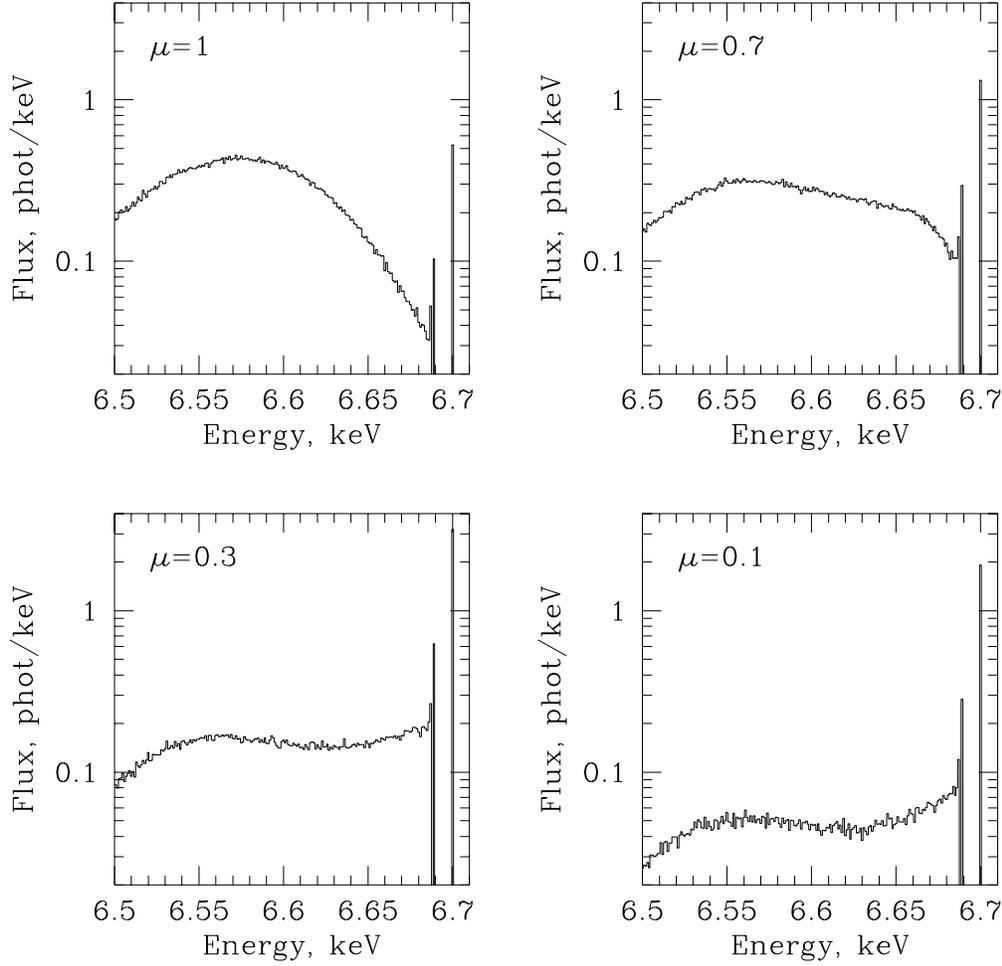}
\caption{
Scattered spectrum of the 6.7 keV line, emitted by the isotropic source above
the plane semi-infinite media with solar abundance of heavy elements, as the
function angle between observer and perpendicular to the surface. $\mu$ is
equal to 
the cosine of this angle. In real conditions the presence of several 
sufficiently broad 
lines will cause additional smearing of the details seen in the figure. 
}
\label{fl67}
\end{figure}

In principle significant differences in the line profiles for the scattering
by hydrogen and helium allows one to make use of these effects for the
determination of helium abundance in scattering media. The Sun's surface can
be the most important application. Huge flux in the emission lines of the
hydrogen-- and helium--like ions of iron during powerful solar flares allows
one to use fully high energy resolution of new X-ray detectors. Shown in
Fig.\ref{fl67} are the typical line profiles, resulting from the reflection
of the isotropic flare above the plane surface (Monte--Carlo simulations,
resonance line of helium-like iron, 6.7 keV). Photoabsorption due to heavy
elements and scattering by atomic hydrogen were taken into account. In real
conditions the presence of blends of relatively broad lines will cause
additional smearing of the features compared to Fig.\ref{fl67}. The most
advantageous conditions for such observations take place when angular
resolution of the telescope allows one to separate the flare itself and the
scattering region.  Of interest are profiles
of the emission lines of hydrogen or helium ions and $K_\alpha$ emission
emerging from the Sun's surface, which is illuminated by the X-ray
continuum. The most interesting feature to look for might be the lines
resulting from Raman scattering by the neutral helium. In comparison with
Fig.\ref{fl67} 
scattering by helium should lead to additional features at the energies
$\sim h\nu_0-20$ eV.  
 
\subsection{$10^4$K plasma in the vicinity of QSOs and active galactic nuclei}
\begin{figure}
\epsfxsize=14cm
\epsffile{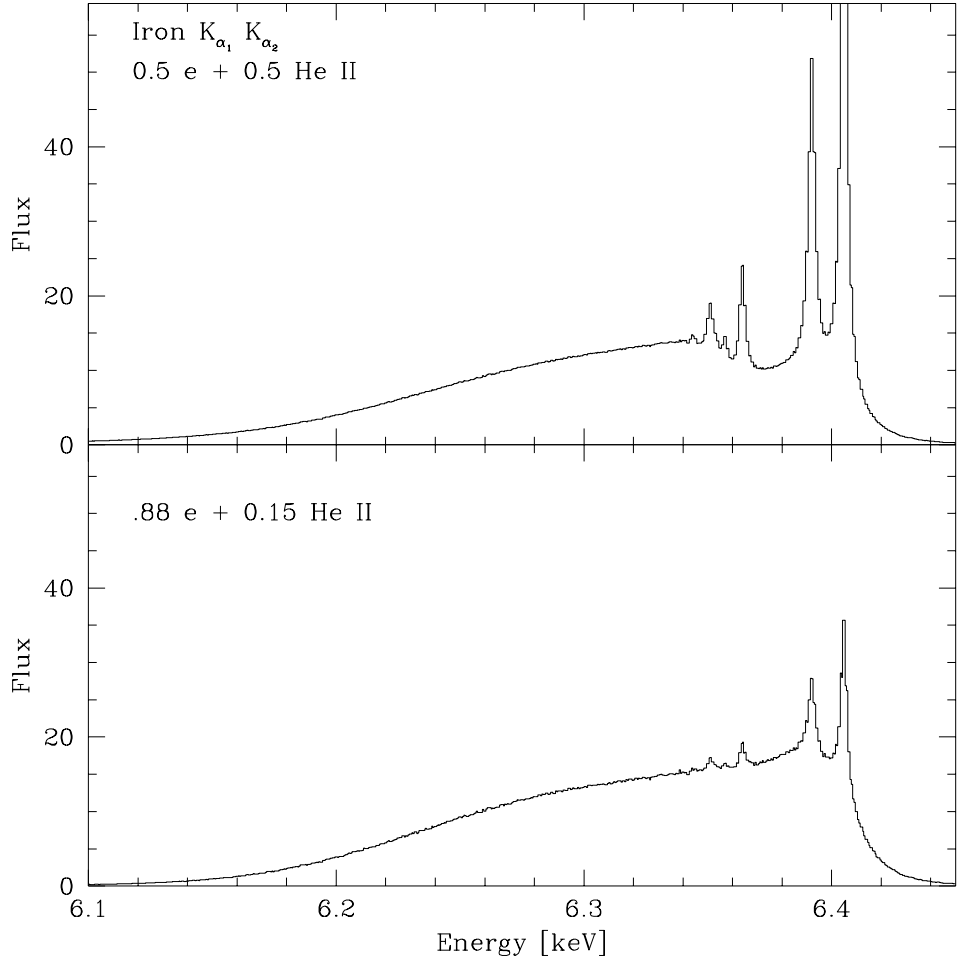}
\caption{
Scattered spectrum of iron fluorescent lines by the mixture of free
electrons (at $T\sim 10^4$ K) and singly ionized helium, averaged over all
angles. For the upper figure fraction of electrons corresponds to the pure
helium 
plasma, for the lower figure fraction of electrons corresponds to the hydrogen
plasma with admixture (15\%) of helium}
\label{hhe}
\end{figure}

In the vicinity of QSOs and active galactic nuclei plasma clouds may exist
with substantial Thomson optical depth, where hydrogen is completely ionized
and helium is present in the form of hydrogen-like ions. This opens the
possibility to observe scattering by the hydrogen-like ions of helium with
the energy gap of 40.8 eV (Fig.\ref{hhe}).

Note finally, that the Raman satellite may also appear due to scattering by the
elements heavier than hydrogen or helium. The major parameters,
determining the strength of the satellites (provided sufficient Thomson
optical depth) is the abundance of the given element and the presence of the
energy 
levels, which excitation energy is comparable with the characteristic energy
change due to recoil effect. From this point of view young supernova
remnants, overabundant with heavy elements may be of special interest.

This work was supported in part by the grants RBRF 96-02-18588-A and INTAS
93-3364. We are grateful to I.Beigman, S.Grebenev, L.Presnyakov,
I.Sobelman, J.Truemper and Y.Tanaka for discussions and important comments.


\clearpage

\pagebreak

\end{document}